\documentclass[aps,pre,reprint,superscriptaddress]{revtex4-1}

\usepackage[dvipdfmx]{graphicx}
\usepackage{amsmath}
\usepackage{dcolumn}
\usepackage{bm}
\usepackage{accents}
\usepackage{amssymb}
\usepackage{mathtools}
\usepackage{braket}
\usepackage{feynmp}
\usepackage{url}
\usepackage{setspace}
\usepackage{color}
\usepackage{pifont}
\usepackage{hyperref}

\newcommand{\revision}[1]{{{#1}}}

\newcommand{\tr}{\mathrm{tr}}

\begin{document}

\title{Thermodynamic approach to quantum cooling limit of continuous Gaussian feedback}

\author{Kousuke Kumasaki}
\affiliation{Department of Applied Physics, The University of Tokyo, 7-3-1 Hongo, Bunkyo-ku, Tokyo 113-8656, Japan}
\email{kumasaki@noneq.t.u-tokyo.ac.jp}

\author{Toshihiro Yada}
\affiliation{Department of Applied Physics, The University of Tokyo, 7-3-1 Hongo, Bunkyo-ku, Tokyo 113-8656, Japan}

\author{Ken Funo}
\affiliation{Department of Applied Physics, The University of Tokyo, 7-3-1 Hongo, Bunkyo-ku, Tokyo 113-8656, Japan}

\author{Takahiro Sagawa}
\affiliation{Department of Applied Physics, The University of Tokyo, 7-3-1 Hongo, Bunkyo-ku, Tokyo 113-8656, Japan}
\affiliation{Quantum-Phase Electronics Center (QPEC), The University of Tokyo, 7-3-1 Hongo, Bunkyo-ku, Tokyo 113-8656, Japan}

\date{\today}

\begin{abstract}
Feedback cooling plays a critical role in stabilizing quantum systems and achieving low temperatures, where a key question is to determine the fundamental thermodynamic limits on cooling performance. We establish a fundamental bound on quantum feedback cooling in Gaussian systems, by deriving a generalized second law of thermodynamics involving the kinetic temperatures of the system and a measure of quantum information flow obtained by continuous measurement.
 In contrast to previously known bounds, the obtained bound can be saturated by experimentally feasible situations using the quantum Kalman filter with a large feedback gain, where the cooling efficiency approaches its maximum. Our theoretical result is numerically demonstrated using parameters from an experiment of levitated nanoparticles. Our theory provides a general framework for understanding the thermodynamic constraints on quantum feedback cooling.
\end{abstract}

\maketitle

\section{Introduction} 
Measurement and feedback are fundamental tools that enable noise suppression, stabilization, optimal control, and adaptive control in diverse applications.
In quantum systems, continuous measurement and feedback based on real-time data have been theoretically studied~\cite{Wiseman_Milburn_2009, Jacobs2006, Wiseman1993, Wiseman_1994_PhysRevA.49.2133, Doherty_and_Jacobs_1999, PhysRevLett.96.043003} and experimentally realized for qubit stabilization~\cite{Vijay_2012, Riste_2013}, error correction~\cite{Livingston_2022}, and cooling of oscillatory systems~\cite{Rossi_2018, Ginseler_2012_PhysRevLett.109.103603, Magrini_2021, Kamba_shimizu_aikawa_2023}.
In nanomechanical resonators~\cite{Rossi_2018} and optically trapped nanoparticles~\cite{Ginseler_2012_PhysRevLett.109.103603, Magrini_2021, Kamba_shimizu_aikawa_2023, doi:10.1126/science.abg3027}, state-based feedback control reduces vibrational energy, enabling cooling close to the ground state.
In Gaussian systems, the quantum Kalman filter~\cite{Belavkin1995} provides optimal filtering that enhances the precision of feedback and has been demonstrated experimentally~\cite{Magrini_2021}.

In feedback cooling, the energy of the system is reduced by extracting work through measurement-based control, which can be interpreted as a ``Maxwell's demon'' setup in the context of thermodynamics of information~\cite{Parrondo_Horowitz_Sagawa_2015}.
By incorporating information terms, fundamental thermodynamic relations, such as the second law and the fluctuation theorem,  have been generalized to systems under measurement and feedback in both classical and quantum regimes~\cite{Sagawa_Ueda_2008_PhysRevLett.100.080403,
Jacobs_2009_PhysRevA.80.012322,Sagawa_Ueda_2010_PhysRevLett.104.090602,
Horowitz_Jordan_Vaikuntanathan_2010_PhysRevE.82.061120,Funo_Watanabe_Ueda_2013_PhysRevE.88.052121, Gong_Ashida_Ueda_PhysRevA.94.012107}, including the case of continuous measurement and feedback~\cite{Sagawa_PRE_2012,Ito_Sagawa_2013_PhysRevLett.111.180603,Horowitz_Esposito_2014_PhysRevX.4.031015,Shiraishi_Sagawa_2015_PhysRevE.91.012130,Horowitz_2015, Krzysztof_Esposito_2019PhysRevLett.122.150603,Yada_2022}.
Experimental studies have further validated those information-thermodynamic relations~\cite{Toyabe_2010, berut2012experimental, Koski_2014, ribezzi2019large, debiossac_2020thermodynamics, debiossac2022non,
Camati_2016,Nathanael_2017_pnas.1704827114, Masuyama_2018, Naghiloo_2018_PhysRevLett.121.030604, Yada_2024}.

A fundamental challenge in feedback cooling is to determine the ultimate limits imposed by thermodynamics. While the standard formulation of the second law relates entropy production to the temperature of the heat bath, the primary objective in feedback cooling is to reduce the temperature of the system below the bath temperature. 
Thermodynamics of information for classical systems have revealed second-law-like inequalities for Langevin systems, which relate the kinetic temperatures of the system itself to the rate of information acquisition through measurement, quantified by transfer entropy~\cite{Horowitz_and_Sandberg_2014, Sandbaerg_2014_PhysRevE.90.042119, Hartich_2014}. However, a quantum counterpart to these results remains elusive, and a general characterization of cooling limits based on thermodynamic laws is still lacking for quantum systems.

\begin{figure}[thbp]
\begin{center}
\includegraphics[width=0.95\linewidth]{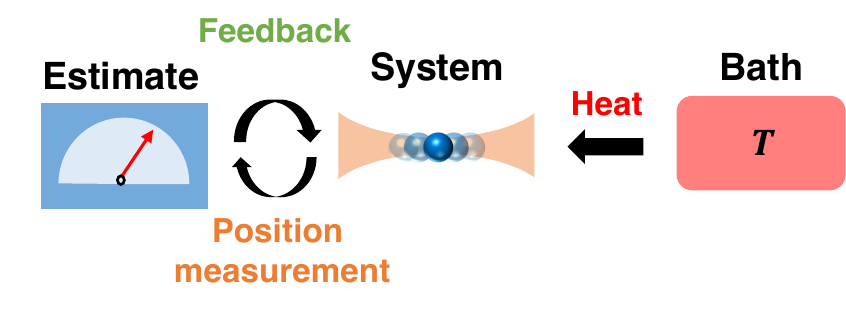}
\caption{Schematic of the setup. The system undergoes feedback control based on estimated values obtained from the \revision{outcomes} of position measurements. The system is also coupled to a heat bath at temperature $T$.}
\label{fig: setup}
\end{center}
\end{figure}

In this study, we consider a setup in which cooling is achieved by a continuous position measurement~\cite{Jacobs2006} together with feedback based on its measurement outcomes~\cite{Wiseman_Milburn_2009, Doherty_and_Jacobs_1999} (see Fig.~\ref{fig: setup}). In this framework, we establish a fundamental bound of feedback cooling in Gaussian quantum systems. 
Specifically, we derive a second-law-like inequality that imposes a constraint on work extraction by a measure of quantum information flow called quantum-classical-transfer (QC-transfer) entropy~\cite{Yada_2022}. 
Since the obtained bound is based on the kinetic temperatures of the system, it enables characterization of feedback cooling (i.e., reduction of the temperature of the system) and is  achievable with a sufficiently large feedback gain along with using the quantum Kalman filter, in contrast to previously known bounds~\cite{Yada_2022}.
 We numerically demonstrate these results with experimentally relevant parameters adopted from a recent nanoparticle experiment~\cite{Magrini_2021}.
 Our results provide a theoretical foundation for the limits of feedback cooling from the perspective of thermodynamics of information in the quantum regime.

This paper is organized as follows. In Sec.~\ref{sec: Setup}, we introduce the setup for Gaussian feedback cooling and define key quantities including the transfer entropy and kinetic temperatures. In Sec.~\ref{sec: main results}, we present the main result, i.e., the information thermodynamic limit of quantum feedback cooling. In Sec.~\ref{ssec: achivement of the cooling limit}, we discuss the asymptotic achievement of the equality condition of the obtained bound.  Finally, in Sec.~\ref{sec: conclusion and outlook}, we summarize our results and outline possible future directions.

\section{Setup}\label{sec: Setup}
\subsection{Model of continuous Gaussian feedback cooling}
We describe the framework for feedback cooling in Gaussian systems, where the state remains Gaussian throughout its evolution.
\revision{We consider continuous position measurement, whose measurement operator commutes with $\hat{x}$, and real-time feedback control based on the measurement record, applied to a Gaussian system coupled to a single heat bath that satisfies detailed balance (see Fig.~\ref{fig: setup}).}
Specifically, the Hamiltonian of the system is supposed to be harmonic: $\hat{H}_{\mathrm{HO}} = \frac{\hbar \omega}{2}\left(\hat{p}^2 + \hat{x}^2\right)$.
Here, $\hat{x}$ and $\hat{p}$ represent the dimensionless position and momentum operators, respectively, satisfying the commutation relation $[\hat{x}, \hat{p}] = i$, and $\omega$ is the angular frequency of the oscillator.
\revision{In addition, the system is externally controlled by a feedback Hamiltonian $\hat{H}_{\mathrm{FB}}(t)$ based on the past measurement outcomes.}
We assume that the expectation values of the initial position and momentum are zero.
The dynamics of feedback cooling are governed by the stochastic master equation~\revision{\cite{Jacobs2006}}:
\begin{align}
\label{eq: nanoparticle stochastic master equation}
d \hat{\rho}_{\mathrm{c}}(t)=&-\frac{i}{\hbar}\left[\hat{H}_{\mathrm{HO}}+\hat{H}_{\mathrm{FB}}(t), \hat{\rho}_{\mathrm{c}}(t)\right]dt+\mathcal{L}_\mathrm{bath}[\hat{\rho}_{\mathrm{c}}(t)]dt \notag \\
&+ 2k\mathcal{D}_{\hat{x}}[\hat{\rho}_{\mathrm{c}}(t)] dt + \sqrt{2k\eta}\mathcal{H}_{\hat{x}}[\hat{\rho}_{\mathrm{c}}(t)] dW,
\end{align}
where 
$\mathcal{D}_{\hat{z}}[\hat{\rho}_{\mathrm{c}}]\coloneqq \hat{z}\hat{\rho}_{\mathrm{c}}\hat{z}^\dagger - \frac{1}{2}\{\hat{z}^\dagger \hat{z},\hat{\rho}_{\mathrm{c}}\}$ is the dissipator and $\hat{\rho}_{\mathrm{c}}(t)$ represents 
\revision{a state conditioned on all measurement outcomes up to time $t$.}
The superoperator $\mathcal{L}_\mathrm{bath}[\hat{\rho}_{\mathrm{c}}]\coloneqq\gamma(\bar{n}+1) \mathcal{D}_{\hat{a}}[\hat{\rho}_{\mathrm{c}}]+\gamma \bar{n}\mathcal{D}_{\hat{a}^{\dagger}}[\hat{\rho}_{\mathrm{c}}]$ represents the contribution of the heat bath, where $\hat{a}$ and $\hat{a}^{\dagger}$ represent the annihilation and creation operators, and $\gamma$ is the dissipation rate.
The ratio between $\bar{n}$ and $\bar{n}+1$ satisfies the detailed balance condition $\bar{n}/(\bar{n}+1)=e^{-\hbar \omega/T}$, where $T$ is the temperature of the heat bath and the Boltzmann constant is set equal to unity for the sake of simplicity.
\revision{The third term on the right-hand side of Eq.~\eqref{eq: nanoparticle stochastic master equation} represents momentum diffusion at a rate $2k$, resulting from the backaction of the position measurement, while the fourth term represents the information obtained from the measurement process~\cite{Jacobs2006}.}
\revision{Because the measurement operator $\sqrt{2k}\hat x$ commutes with $\hat{x}$, the measurement is a quantum-non-demolition (QND) position measurement if the Hamiltonian term $\hat{H}_{\mathrm{HO}}$ is absent.}
The superoperator $\mathcal{H}_{x}$ is defined as $\mathcal{H}_{\hat{x}}[\hat{\rho}_{\mathrm{c}}]\coloneqq \hat{x} \hat{\rho}_{\mathrm{c}} + \hat{\rho}_{\mathrm{c}}\hat{x}  -2\mathrm{tr}(\hat{x}\hat{\rho}_{\mathrm{c}})\hat{\rho}_{\mathrm{c}}$.
The parameter $k$ represents the measurement rate, while the Wiener increment $dW$ is a Gaussian random variable with zero mean and variance $dt$.
The measurement \revision{outcome} obtained via measurement of the position over \revision{an infinitesimal time interval} $dt$ is described by $d y(t)=\braket{\hat{x}(t)}_{\mathrm{c}}dt+dW/\sqrt{8k\eta}$, where $\eta$ represents the detection efficiency and $\braket{\cdot(t)}_{\rm{c}}\coloneqq\mathrm{tr}\left(\hat{\rho}_{\rm{c}}(t)\;\cdot\right)$ denotes the expectation value under the conditional state.
The set of all measurement \revision{outcomes} until time $t$ is defined as $Y_{t} \coloneqq \{dy(s)\}_{s=0}^{t^{-}}$.
\revision{Note that $\hat{\rho}_{\mathrm{c}}(t)$ is the state of the system for a given realization of measurement outcomes $Y_t$. We also denote $\hat \rho = \mathbb{E}_{Y_t} [\hat \rho_{\rm c}]$ as the averaged state by taking average over all measurement outcomes, where $\mathbb{E}_{X}$ represents taking the expectation value over a random variable $X$.}

Estimates of the observable are constructed from the set of obtained measurement outcomes and used for feedback.
Using measurement \revision{outcomes} $Y_{t}$, estimates are defined as $e_{t}=h(Y_{t})$, where $h(\cdot)$ compresses the past measurement data into estimates $e_{t}$.
The posterior state conditioned on this estimate is denoted by $\hat{\rho}_{t}^{e_{t}}$.
Using the estimates, the feedback action is determined.
The linear feedback Hamiltonian $\hat{H}_{\rm{FB}}$ is given by
\begin{equation}
\hat{H}_{\mathrm{FB}}(t)= \hbar c_x(e_t) \hat{x}+\hbar c_p(e_t) \hat{p}, \label{eq: feedback hamiltonian}
\end{equation}
where $c_x$ and $c_p$ are linear functions of the estimate $e_t$, defining the feedback coefficients.
The special case of the estimate is the quantum Kalman filter~\cite{Belavkin1995}, which is given by the conditional expectation values
\begin{equation}\label{eq: quantum Kalman filter}
   e_t = (\braket{\hat{x}(t)}_{\mathrm{c}},\braket{\hat{p}(t)}_{\mathrm{c}} ). 
\end{equation}
\revision{Note that obtaining the measurement outcomes reduces the uncertainty in position, although $\braket{\hat{x}}_{\mathrm{c}}$ and $\braket{\hat{p}}_{\mathrm{c}}$ fluctuates due to random measurement outcomes. The feedback Hamiltonian is then designed to decrease the absolute values of $\braket{\hat{x}}_{\mathrm{c}}$ and $\braket{\hat{p}}_{\mathrm{c}}$, and thus the fluctuation of position and momentum is reduced at the ensemble average level. Therefore, the dynamics described by Eq.~(\ref{eq: nanoparticle stochastic master equation}) allows reducing the energy of the system and realizes feedback cooling.}

\subsection{QC-transfer entropy} 
As our goal is to obtain an information-thermodynamic relation to characterize the cooling limit of Gaussian systems,  we consider the QC-transfer entropy introduced in Ref.~\cite{Yada_2022}, which quantifies the amount of information gain through continuous measurement.
It is defined as
\begin{align} \dot{I}_{\mathrm{QCT}} \coloneqq \lim_{dt\rightarrow 0}\frac{1}{dt}\mathbb{E}_{Y_t}\left[ I_{\mathrm{QC}}\left(\hat{\rho}_{\mathrm{c}}: dy(t)\right)\right].\label{eq: QC transfer entropy}
\end{align}
Here, $I_{\mathrm{QC}}\left(\hat{\rho}_{\mathrm{c}}: dy(t)\right)$ is  the quantum-classical-mutual (QC-mutual) information~\cite{groenewold_1971,ozawa_1986} given by
\begin{align} \label{eq: QC mutual information}
I_{\mathrm{QC}}\left(\hat{\rho}_{\mathrm{c}}:  dy(t) \right) \coloneqq S\left(\hat{\rho}_{\mathrm{c}}\right) - \mathbb{E}_
{dy(t)}\left[S\left(\hat{\rho}_{\mathrm{c,\mathrm{M}}}\right)\right],
\end{align}
where $\hat{\rho}_{\mathrm{c,\mathrm{M}}}:=\hat{\rho}_{\mathrm{c}}+ 2k\mathcal{D}_{\hat{x}}[\hat{\rho}_{\mathrm{c}}] dt +4k\eta \mathcal{H}_{\hat{x}}[\hat{\rho}_{\mathrm{c}}] (dy(t)-\braket{\hat{x}(t)}_{\mathrm{c}}dt)$ is the conditional state after measurement at time $t$, and $S(\hat \rho) \coloneqq-\mathrm{tr}[\hat \rho \ln \hat \rho]$ represents the von Neumann entropy.
We note that Eq.~\eqref{eq: QC mutual information} is regarded as the conditional QC-mutual information given all prior measurement \revision{outcomes} $Y_t$, because $\hat \rho_{\rm c}$ is the conditional state.
Since the QC-transfer entropy quantifies the total information gain, it provides an upper bound for the information available in the multiple feedback processes.

\begin{figure}[tbp]
\begin{center}
\includegraphics[width=0.95\linewidth]{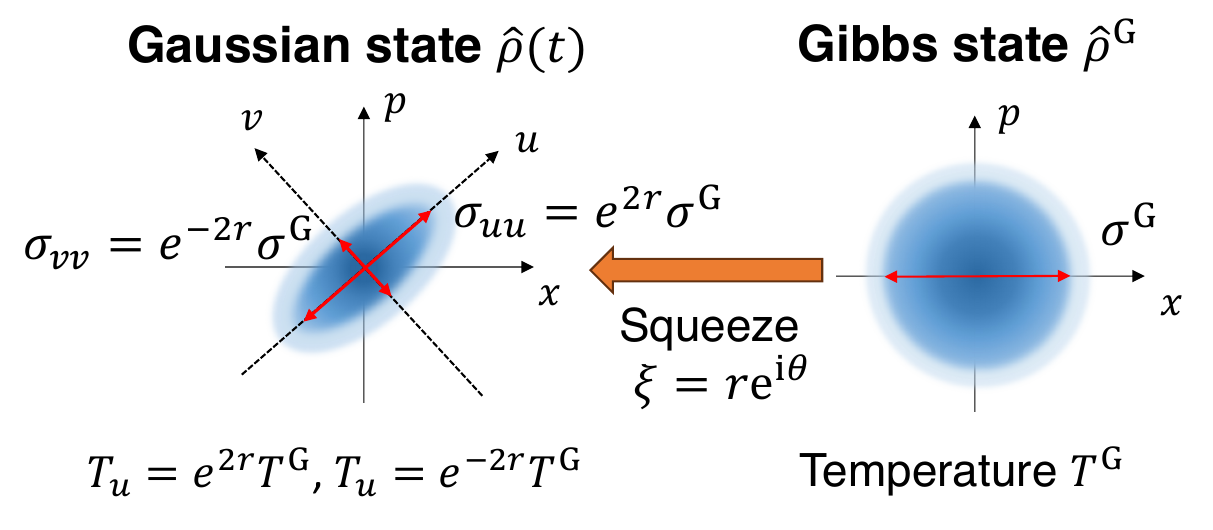}
\caption{Schematic of the distribution of a Gaussian state $\hat{\rho}(t)$ in the phase space (i.e., the Wigner distribution).  The kinetic temperatures $T_u$ and $T_v$ characterize the variances in the eigen-directions of the covariance matrix $\boldsymbol{\sigma}$. 
}
\label{fig: gaussian state}
\end{center}
\end{figure}

\subsection{Kinetic temperatures}
We here introduce kinetic temperatures and work rate.
We begin by defining the kinetic temperatures of a given Gaussian state.
The one-mode Gaussian state is fully characterized by the mean values of position and momentum and the variance-covariance matrix.
Let $\boldsymbol{\sigma}:=[\sigma_{ij}]_{i,j = x,p}$ denote the covariance matrix of the averaged state $\hat{\rho}(t)$.
The position and momentum variances are given by $\sigma_{xx} \coloneqq \braket{\hat{x}^2} - \braket{\hat{x}}^2$ and $\sigma_{pp} \coloneqq \braket{\hat{p}^2} - \braket{\hat{p}}^2$, respectively, while the covariance is defined as $\sigma_{xp} \coloneqq \frac{1}{2} \braket{\{\hat{x}, \hat{p}\}} - \braket{\hat{x}}\braket{\hat{p}}$, where $\langle \cdot \rangle$ denotes the expectation value with respect to $\hat{\rho}(t)$.
Let the eigenvalues of $\boldsymbol{\sigma}$ be $\sigma_{uu}$ and $\sigma_{vv}$.
Noting that a single-mode Gaussian state can always be expressed as a squeezed thermal state~\cite{Kim_1989}, we define the kinetic temperatures along eigen-directions.
The single-mode squeezed thermal state is generated by applying a squeezing operation $\hat{S}(\xi) := \exp \left( \frac{1}{2}\xi (\hat{a}^{\dagger })^2 -\frac{1}{2}\xi^* \hat{a}^2    \right)$ with a squeezing parameter $\xi=r\rm{e}^{i\theta}$ to the Gibbs state $\hat{\rho}^{\mathrm{G}}\coloneqq \exp(-\hat{H}_{\mathrm{HO}}/T^{\mathrm{G}})/\mathrm{tr}[\exp(-\hat{H}_{\mathrm{HO}}/T^{\mathrm{G}})]$~\cite{Olivares_2012}, where $T^{\mathrm{G}}$ is the temperature of the Gibbs state, as $\hat{\rho}_{t} = \hat{S}(\xi) \hat{\rho}^{\mathrm{G}} \hat{S}(\xi)^{\dagger}$.
The eigenvalues of $\boldsymbol{\sigma}$ satisfy $\sigma_{uu} = {\rm{e}}^{2r}\sigma^{\rm{G}}$ and $\sigma_{vv} = {\rm{e}}^{-2r}\sigma^{\rm{G}}$, where 
\begin{align}
    \sigma^{\mathrm{G}}:=\mathrm{tr}(\hat{x}^{2}\hat{\rho}^{\rm G})=\mathrm{tr}(\hat{p}^{2}\hat{\rho}^{\rm G})=\frac{1}{2}\coth \frac{\hbar\omega\beta^{\rm G}}{2},
\end{align}
is the common position and momentum variance of the state $\hat{\rho}^{\mathrm{G}}$.
Accordingly, the kinetic temperatures $T_u$ and $T_v$, associated with these eigenvalues, are defined as:
\begin{align}\label{eq: kinetic temperature}
     T_u \coloneqq {\rm{e}}^{2r}T^{\mathrm{G}}, \quad T_v \coloneqq {\rm{e}}^{-2r}T^{\mathrm{G}}.
\end{align}
A schematic of kinetic temperatures is shown in Fig.~\ref{fig: gaussian state}.
The kinetic temperatures $T_u$ and $T_v$ correspond to scaling the Gibbs state temperature 
\begin{align} \label{eqref: gibbs state eigenvalue relation}
T^{\mathrm{G}} = \hbar \omega \left[ \ln \left( \frac{2\sqrt{\det \boldsymbol{\sigma}} + 1}{2\sqrt{\det \boldsymbol{\sigma}} - 1} \right)\right]^{-1},
\end{align}
according to the widths in eigen-directions of the distribution.
We further note that 
\begin{align}\label{eq: lnrhoTuTv}
    \ln \hat{\rho}_{t} &= \hat{S}(\xi) \left( \ln  \hat{\rho}^{\mathrm{G}} \right) \hat{S}(\xi)^{\dagger} \nonumber  \\
    &= -\beta^{\rm G} \hat{S}(\xi) \hat{H}_{\rm HO} \hat{S}(\xi)^{\dagger} -\ln Z(\beta^{\mathrm{G}}) \nonumber \\
    &=-\frac{\hbar\omega \hat{u}^{2}}{2T_{u}} - \frac{\hbar\omega\hat{v}^{2}}{2T_{v}} -\ln Z(\beta^{\mathrm{G}}),
\end{align}
where 
\begin{align}
    \hat{u}&\coloneqq e^{r}\hat{S}(\xi)\hat{x}\hat{S}(\xi)^{\dagger}, \label{ss:u} \\
    \hat{v}& \coloneqq e^{-r}\hat{S}(\xi)\hat{p}\hat{S}(\xi)^{\dagger}, \label{ss:v}
\end{align}
are the quadratures that correspond to the eigenvalues $\sigma_{uu}$ and $\sigma_{vv}$, i.e., $\sigma_{uu}=\mathrm{tr}(\hat{u}^{2}\hat{\rho}_{t})-\mathrm{tr}(\hat{u}\hat{\rho}_{t})^{2}$ and  $\sigma_{vv}=\mathrm{tr}(\hat{v}^{2}\hat{\rho}_{t})-\mathrm{tr}(\hat{v}\hat{\rho}_{t})^{2}$, respectively. 

We next define the work rate as follows. 
First, the internal energy of the system is given by $E(t)=\mathrm{tr}[\hat{H}_{\mathrm{HO}}\hat{\rho}(t)]$.
Then, based on the first law of thermodynamics, the work rate corresponding to the energy extracted from the system is defined as $\dot{W}_{\mathrm{ext}} := \dot{Q} - \frac{d}{dt}E(t)$, where 
\begin{align}\label{eq: def of heat}\dot{Q}:=\mathrm{tr}\left(\hat{H}_{\mathrm{HO}} \mathcal{L}_{\mathrm{bath}}[\rho]\right),
\end{align}is the heat~\cite{Gong_Ashida_Ueda_PhysRevA.94.012107}.
The time variation of the energy can be expressed as contributions along the directions of $\sigma_{uu}$ and $\sigma_{vv}$, using the quadrature operators $\hat{u}$ and $\hat{v}$ associated with these directions, as $\frac{d}{dt}E(t) = \frac{\hbar\omega}{2} \frac{d}{dt}\mathrm{tr}\left[\hat{\rho} \hat{u}^2\right] + \frac{\hbar\omega}{2} \frac{d}{dt}\mathrm{tr}\left[\hat{\rho} \hat{v}^2\right]$.
Similarly, the work rate can be further decomposed as:
\begin{align}
\dot{W}_{\mathrm{ext}} = \dot{W}_{\mathrm{ext},u} + \dot{W}_{\mathrm{ext},v},
\end{align}
where the terms $\dot{W}_{\mathrm{ext},u}$ and $\dot{W}_{\mathrm{ext},v}$ represent the contributions of the work rate along the directions of $\sigma_{uu}$ and $\sigma_{vv}$, defined by Eqs.~\eqref{eq: Wextu} and \eqref{eq: Wextv}.

\section{Main results}\label{sec: main results}
\subsection{Information thermodynamic cooling limit}
We now present the main result of this study, a fundamental bound of the work rate in terms of kinetic temperatures.
Suppose that the system is in the steady state; the \revision{averaged} state $\hat \rho$ is given by the steady solution of Eq.~\eqref{eq: nanoparticle stochastic master equation}.
Then, we have a generalized second law of thermodynamics in the form
\begin{align}\label{eq: steady state inequality} 
\frac{\dot{W}_{\mathrm{ext},u}}{T_u} + \frac{\dot{W}_{\mathrm{ext},v}}{T_v} \leq \dot{I}_{\mathrm{QCT}}. 
\end{align}
This inequality indicates how much QC-transfer entropy can be utilized to enhance work extraction and lower the kinetic temperatures, and serves as a quantum generalization of the second-law-like bound for classical Langevin systems~\cite{Horowitz_and_Sandberg_2014}. Since cooling is achieved by extracting work from the system and lowering its kinetic temperatures, inequality~\eqref{eq: steady state inequality}  represents the fundamental cooling limit.

We compare our result~\eqref{eq: steady state inequality} with the generalized second law using the temperature $T$ of the heat bath, given in the form~\cite{Yada_2022}: 
\begin{align}\label{eq: Usual SL}
\frac{\dot{W}_{\rm{ext}}}{T}\leq \dot{I}_{\mathrm{QCT}}.
\end{align} 
In Sec.~\ref{sec: comparison witha the second law} of the appendix, we show that the following relation holds in the steady state: 
\begin{align}\label{eq: second law and cooling limit}
\frac{\dot{W}_{\mathrm{ext}}}{T}
\le
\frac{\dot{W}_{\mathrm{ext},u}}{T_u}
+
\frac{\dot{W}_{\mathrm{ext},v}}{T_v}.
\end{align}
This relation shows that inequality~\eqref{eq: steady state inequality} provides a tighter bound compared to~\eqref{eq: Usual SL} in the steady-state regime, since the former explicitly involves the kinetic temperatures of the system itself.

We next consider the conditions for achieving the equality in \eqref{eq: steady state inequality}.
A key ingredient is the quantum Kalman filter, which provides the conditional expectation values as the estimates, as presented in Eq.~\eqref{eq: quantum Kalman filter}.
Note that $\hat{\rho}_{t}^{e_{t}} = \hat{\rho}_{\mathrm{c}}$ holds in this case.
By using this estimate based on the quantum Kalman filter, one can implement the so-called state-based feedback~\cite{Doherty_and_Jacobs_1999, Magrini_2021}, where the feedback Hamiltonian~\eqref{eq: feedback hamiltonian} is 
given by the coefficients $c_x=b_x \langle \hat{x}\rangle_{\mathrm{c}}+b_p\langle \hat{p}\rangle_{\mathrm{c}}$ and $c_p=-a_x \langle \hat{x}\rangle_{\mathrm{c}}-a_p\langle \hat{p}\rangle_{\mathrm{c}}$. 
This feedback Hamiltonian aims to reduce the mean position and momentum to the origin in the phase space.
Another requirement for achieving the equality in \eqref{eq: steady state inequality} is a large feedback gain.
In fact, the equality in~\eqref{eq: steady state inequality} is asymptotically achieved  as the feedback gain goes to infinity if the estimate is implemented by the quantum Kalman filter (see Sec.~\ref{ssec: achivement of the cooling limit} for the proof).
For simplicity, let the parameters of the feedback Hamiltonian be $a_x = b_p = g$ and $a_p = b_x = 0$.
In this case, the equality in~\eqref{eq: steady state inequality} is asymptotically achieved in the limit $g \to \infty$.
This implies that, under these conditions, the feedback system perfectly converts the acquired information into work.

Under the same conditions, the occupancy number of the harmonic Hamiltonian $\hat H_{\rm HO}$, $\langle a^{\dagger}a\rangle$, also reaches its minimal value determined by the bath temperature and the measurement rate (see Sec.~\ref{ssec: achivement of the cooling limit}), as the optimal feedback minimizes the sum of the variances of position and momentum, which is equivalent to the occupation number in our setup.
This highlights the connection between the information-theoretic bound on thermodynamic energy conversion and the cooling limit of Gaussian systems. 


\begin{figure}[tbp]
\begin{center}
\includegraphics[width=0.95\linewidth]{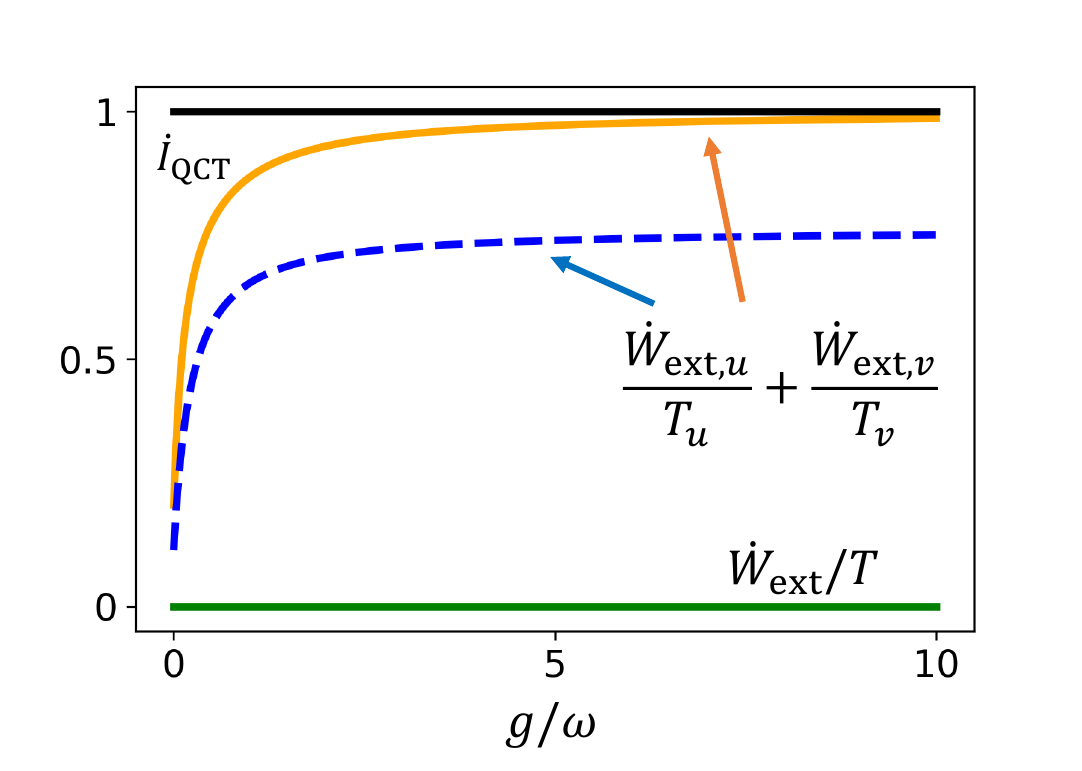}
\caption{The behavior of ${\dot{W}_{\mathrm{ext},u}}/{T_u} + {\dot{W}_{\mathrm{ext},v}}/{T_v}$ as a function of the feedback gain $g$ in the steady state of Eq.~\eqref{eq: nanoparticle stochastic master equation}.
The feedback protocol is based on the quantum Kalman filter, where the feedback Hamiltonian is given by $H^{\hat{x},\hat{p}}_{\rm{FB}} =\hbar g(-\langle x \rangle_{\rm{c}} \hat{p} + \langle p \rangle_{\rm{c}} \hat{x})$ (orange curve).
In addition, we demonstrate another feedback Hamiltonian $H^{\hat{x}}_{\rm{FB}} =\hbar g(\langle x \rangle_{\rm{c}} + \langle p \rangle_{\rm{c}})\hat{x}$ (blue dashed curve) for the sake of comparison.
The behavior of ${\dot{W}_{\rm{ext}}}/{T}$ is also shown (green curve).
The black line represents the QC-transfer entropy, the bound of~\eqref{eq: steady state inequality}, to which only the orange curve approaches in large $g$.
The parameters are chosen as $k/\omega = 0.18$, $\eta=0.34$, $T=292$K, and $\bar{n}\gamma/\omega =  0.0058$~\cite{Magrini_2021}.  \revision{Note that the kinetic temperatures $T_{u}$ and $T_{v}$ are orders of magnitude smaller than the bath temperature, and thus the green line becomes approximately equal to zero.} The vertical axis is normalized by $\dot{I}_{\mathrm{QCT}}$.
}
\label{fig: changing feedback gain}
\end{center}
\end{figure}

\subsection{Numerical demonstration}
We present a numerical demonstration of our main inequality~\eqref{eq: steady state inequality}, by numerically implementing the quantum Kalman filter.
Figure~\ref{fig: changing feedback gain} shows the behavior of ${\dot{W}_{\mathrm{ext},u}}/{T_u} + {\dot{W}_{\mathrm{ext},v}}/{T_v}$ and ${\dot{W}_{\rm{ext}}}/{T}$ as functions of the feedback gain in the steady state of Eq.~\eqref{eq: nanoparticle stochastic master equation}.
The parameters are taken from Ref.~\cite{Magrini_2021}, an experiment of feedback cooling of levitated nanoparticles.

Figure~\ref{fig: changing feedback gain} compares the two feedback Hamiltonians and shows that only $H^{\hat{x},\hat{p}}_{\rm{FB}} = \hbar g(-\langle x \rangle_{\rm{c}} \hat{p} + \langle p \rangle_{\rm{c}} \hat{x})$, plotted in the orange curve, ensures that ${\dot{W}_{\mathrm{ext},u}}/{T_u} + {\dot{W}_{\mathrm{ext},v}}/{T_v}$ approaches the upper bound $\dot{I}_{\mathrm{QCT}}$, as $g \to \infty$.
Since feedback proportional to $\hat{x}$ is commonly used in experiments, we also consider the Hamiltonian of the form $H^{\hat{x}}_{\rm{FB}} = \hbar g(\langle x \rangle_{\rm{c}} + \langle p \rangle_{\rm{c}})\hat{x}$ (blue dashed curve),
but it does not allow the system to reach the information-theoretic bound of~\eqref{eq: steady state inequality}.
Moreover, $\dot{W}_{\rm{ext}}/T$ with feedback Hamiltonian $H^{\hat{x},\hat{p}}_{\rm{FB}}$ (green curve) fails to asymptotically approach $\dot{I}_{\rm QCT}$, indicating that the naive bound~\eqref{eq: Usual SL} without using the kinetic temperatures is not achievable.

\revision{In addition, we compared our measurement scheme with an alternative scheme that produces the same heating rate for the occupation number. We performed numerical simulations for both methods and demonstrated that our protocol has an advantage in cooling (see Sec.~\ref{sec: homodyne} of the appendix).}

\subsection{Outline of the proof, and more on quantum information flow}\label{sec: proof of main result}
To outline the derivation of our main inequality~\eqref{eq: steady state inequality} and to clarify the structure behind its equality condition, here we introduce yet another information quantity, named the quantum-classical (QC) information flow, which is slightly different from the QC-transfer entropy.
The QC information flow is defined as
\begin{equation}\label{eq: infomation flow s}
    \dot{I}_{\mathrm{QCI}}^{\mathrm{S}} \coloneqq \lim_{dt \rightarrow 0}\frac{1}{dt}\biggl[ \chi (\hat{\rho}_{t+dt} : e_{t}) - \chi (\hat{\rho}_{t} : e_{t})\biggr],
\end{equation}
where $\chi (\rho:e) \coloneqq S(\rho) - \mathbb{E}_e\left[S(\rho^e)\right]$ is the Holevo information~\cite{holevo_1973}.
Since the Holevo information is the measure of the correlation between a quantum state and a classical memory, the QC information flow $-\dot{I}_{\mathrm{QCI}}^{\mathrm{S}}$ quantifies the change in the system-memory correlation induced by the time evolution of the system during the feedback.
We note that the QC information flow is a quantum version of the classical information flow \cite{Horowitz_Esposito_2014_PhysRevX.4.031015}, which is defined as the partial derivative of the classical mutual information.

Based on the QC information flow, we can derive a generalized second law that is also applicable to non-steady states:
\begin{align}\label{eq: inequality for QC information flow and work rate}
\frac{\dot{W}_{\mathrm{ext},u}}{T_u} + \frac{\dot{W}_{\mathrm{ext},v}}{T_v}\leq-\dot{I}_{\mathrm{QCI}}^S - \dot{S}_{\mathrm{BA}}.
\end{align}
Here, $\dot{S}_{\mathrm{BA}}$ represents the entropy change due to measurement backaction, defined as
\begin{align}
 \dot{S}_{\mathrm{BA}}:=\lim_{dt \rightarrow 0}\frac{1}{dt}\mathbb{E}_{e_t}\left[\left(S\left(\hat{\rho}_{t,\mathrm{BA}}^{e_{t}}\right)-S\left(\hat{\rho}_{t}^{e_{t}}\right)\right)\right],
 \label{eq: backaction entropy change}
\end{align}
where $\hat{\rho}^{e_t}_{t,\mathrm{BA}}:=\hat{\rho}_{t}^{e_{t}}+ 2k\mathcal{D}_{\hat{x}}[\hat{\rho}_{t}^{e_{t}}] dt$ is the state evolved for a time interval $dt$ under the contributions of measurement backaction.
The term $-\dot{I}_{\mathrm{QCI}}^{\mathrm{S}} - \dot{S}_{\mathrm{BA}}$ quantifies the information available for feedback control, because the unavoidable entropy change due to the measurement backaction is subtracted from the change in correlation originating from the dynamics of the system.
The detailed derivation of \eqref{eq: inequality for QC information flow and work rate} is provided in Sec.~\ref{sec: Info-thermo cooling limit derivation} of the appendix.
A sufficient condition to achieve the equality of~\eqref{eq: inequality for QC information flow and work rate} is the quantum Kalman filter and a large feedback gain.

We next relate the two measures of quantum information flow: QC-transfer entropy and QC information flow.
Again suppose that the system is in the steady state of Eq.~\eqref{eq: nanoparticle stochastic master equation}.
Then, the QC-transfer entropy provides an upper bound of the QC information flow:
\begin{align}\label{eq: inequality for information flow and transfer entropy}
-\dot{I}_{\mathrm{QCI}}^{\mathrm{S}} -\dot{S}_{\mathrm{BA}}\leq \dot{I}_{\mathrm{QCT}}.
\end{align}
See Sec.~\ref{sec: QCT QCI derivation} of the appendix for the derivation.
A sufficient condition for the equality in~\eqref{eq: inequality for information flow and transfer entropy} is the quantum Kalman filter.
For the classical case of~\eqref{eq: inequality for information flow and transfer entropy}, the equality condition can be expressed in terms of sufficient statistics~\cite{Matsumoto_2018}.

Now, by combining inequalities~\eqref{eq: inequality for QC information flow and work rate} and \eqref{eq: inequality for information flow and transfer entropy}, 
our main result \eqref{eq: steady state inequality} is obtained.  The equality condition for \eqref{eq: steady state inequality} directly follows from the equality conditions for inequalities~\eqref{eq: inequality for QC information flow and work rate} and \eqref{eq: inequality for information flow and transfer entropy}.

\section{Achievement of the cooling limit based on the quantum Kalman filter}\label{ssec: achivement of the cooling limit}
An example of experimental systems that obey the stochastic master equation \eqref {eq: nanoparticle stochastic master equation} is the levitated nanoparticle system~\cite{Magrini_2021}.
In nanoparticle systems, the degree of cooling is often evaluated by how small the occupation number of the particle, $\langle n \rangle = \langle a^\dagger a \rangle$, can be made.
Using the variances of the state, this can be expressed as $\braket{n}=\frac{1}{2}\left(\sigma_{xx} + \sigma_{pp}\right) - 1/2$ .
Defining the occupation number corresponding to  $\boldsymbol{\sigma}_{\mathrm{c}}$ as $\langle n \rangle_{\min}\coloneqq\frac{1}{2}(\sigma_{xx,\mathrm{c}}+\sigma_{pp,\mathrm{c}})$, we obtain:
\begin{align}\label{eq: cooling limit}
\braket{n}\geq\braket{n}_{\mathrm{min}}:=\frac{1}{2}\left(\sigma_{xx,\mathrm{c}}^2 + \sigma_{pp,\mathrm{c}}^2\right) - \frac{1}{2}.
\end{align}
Here, $\langle n \rangle_{\min}$ represents the cooling limit.
The equation of the inequality~\eqref{eq: cooling limit} is fulfilled when $\boldsymbol{\sigma} = \boldsymbol{\sigma}_{\mathrm{c}}$ holds.
In the absence of a heat bath, the equality condition for~\eqref{eq: cooling limit} is known to be satisfied in the limit of strong feedback using the quantum Kalman filter estimates~\cite{Doherty_and_Jacobs_1999}.
A similar sufficient condition can also be demonstrated in this case.
At the same time, the equality condition of the inequality (\ref{eq: steady state inequality}) holds.
This means that the information thermodynamic bound obtained in this study and the cooling limit of nanoparticle systems are reached simultaneously.
In what follows, we show the equality conditions of Eqs.~(\ref{eq: cooling limit}) and (\ref{eq: steady state inequality}). 

We now briefly introduce the quantum Kalman filter, in which the estimated value $e_t=\{\braket{\hat{x}}_{\mathrm{c}},\braket{\hat{p}}_{\mathrm{c}}\}$ is obtained. 
We assume that a feedback Hamiltonian $\hat{H}_{\mathrm{FB}}(t)=-\left(a_x \braket{x}_{\rm{c}}+a_p\braket{p}_{\rm{c}}\right) \hat{p}+\left(b_x \braket{x}_{\rm{c}}+b_p\braket{p}_{\rm{c}}\right) \hat{x}$ is applied using this estimated value.
Then, time evolutions of the expected values $\braket{\hat{x}}_{\mathrm{c}}$ and  $\braket{\hat{p}}_{\mathrm{c}}$ governed by Eq.~\eqref{eq: nanoparticle stochastic master equation} are expressed as:
\begin{align}
d\braket{\hat{x}}_{\mathrm{c}}=&\left(\revision{\omega}\braket{\hat{p}}_{\mathrm{c}}-a_x\braket{\hat{x}}_{\mathrm{c}}-a_p\braket{\hat{p}}_{\mathrm{c}}-\frac{\gamma}{2}\braket{\hat{x}}_{\mathrm{c}}\right)dt \nonumber \\
&+ 8k\eta\sigma_{xx,\mathrm{c}}\cdot (dy - \braket{x}_{\mathrm{c}}dt), \label{eq: ito differential equation xc}\\
d\braket{\hat{p}}_{\mathrm{c}}=&\left(-\omega\braket{\hat{x}}_{\mathrm{c}}-b_x\braket{\hat{x}}_{\mathrm{c}}-b_p\braket{\hat{p}}_{\mathrm{c}}-\frac{\gamma}{2}\braket{\hat{p}}_{\mathrm{c}}\right)dt \nonumber \\
&+ 8k\eta\sigma_{xp,\mathrm{c}}\cdot (dy - \braket{x}_{\mathrm{c}}dt). \label{eq: ito differential equation pc}
\end{align}
This stochastic differential equation allows us to determine the expected values of the position and momentum at the next time step based on current expected values and the measurement outcome $dy$.
In the quantum Kalman filter, this mechanism is used to sequentially obtain the estimate $e_t=\{\braket{\hat{x}}_{\mathrm{c}},\braket{\hat{p}}_{\mathrm{c}}\}$.
Here, the variance and covariance of the posterior state appearing in the filtering coefficients can be determined in advance because the differential equations of $\boldsymbol{\sigma}_{\mathrm{c}}$ are closed with respect to $\boldsymbol{\sigma}_{\mathrm{c}}$ [see Eqs.~(\ref{eq: dynamics of sigma xc})-(\ref{eq: dynamics of sigma xpc})]. 

We further analyze the fluctuations inherent in the estimates of the quantum Kalman filter $e_t=\{\braket{\hat{x}}_{\mathrm{c}},\braket{\hat{p}}_{\mathrm{c}}\}$.
The variance and covariance of the estimates are defined as $\sigma_{xx, {\mathrm{m}}} \coloneqq \mathbb{E}_{e_t}[\braket{\hat{x}}_{\mathrm{c}}^2] - (\mathbb{E}_{e_t}[\braket{\hat{x}}_{\mathrm{c}}])^2$, $\sigma_{pp, {\mathrm{m}}} \coloneqq \mathbb{E}_{e_t}[\braket{\hat{p}}_{\mathrm{c}}^2] - (\mathbb{E}_{e_t}[\braket{\hat{p}}_{\mathrm{c}}])^2$, $ 
\sigma_{xp, {\mathrm{m}}} \coloneqq \mathbb{E}_{e_t}[\braket{\hat{x}}_{\mathrm{c}} \braket{\hat{p}}_{\mathrm{c}}] - \mathbb{E}_{e_t}[\braket{\hat{x}}_{\mathrm{c}}] \mathbb{E}_{e_t}[\braket{\hat{p}}_{\mathrm{c}}]$.
We define the covariance matrix of these as $\boldsymbol{\sigma}_{\mathrm{m}}$.
The covariance matrix can be decomposed as $\boldsymbol{\sigma}=\boldsymbol{\sigma}_{\mathrm{c}}+\boldsymbol{\sigma}_{\mathrm{m}}$.
From Eqs.~\eqref{eq: ito differential equation xc} and \eqref{eq: ito differential equation pc},  the differential equations governing the time evolution of $\boldsymbol{\sigma}_{\mathrm{m}}$ are obtained as:
\begin{align}
\frac{d}{d t} \sigma_{xx,{\mathrm{m}}}=&2\omega \sigma_{xp,{\mathrm{m}}}-2 a_x \sigma_{xx,{\mathrm{m}}}-2 a_p\sigma_{xp,{\mathrm{m}}}\nonumber \\
&-\gamma \sigma_{xx,{\mathrm{m}}}+8 k\eta \sigma_{xx,\mathrm{c}}^2\label{eq: daynamics of sigma xm} \\
\frac{d}{d t} \sigma_{pp,{\mathrm{m}}}=&-2 \omega \sigma_{xp,{\mathrm{m}}}-2 b_x \sigma_{xp,{\mathrm{m}}}-2 b_p \sigma_{pp,{\mathrm{m}}}\nonumber \\
&-\gamma \sigma_{pp,{\mathrm{m}}}+8 k \eta \sigma_{xp,{\mathrm{c}}}^2  \label{eq: daynamics of sigma pm}\\
\frac{d}{d t} \sigma_{xp,{\mathrm{m}}}=&- (\omega+b_x) \sigma_{xx,{\mathrm{m}}}+(\omega-a_p)\sigma_{pp,{\mathrm{m}}}\nonumber \\
&-(a_x+b_p+\gamma) \sigma_{xp,{\mathrm{m}}}+8 k \eta \sigma_{xx,\mathrm{c}} \sigma_{xp,\mathrm{c}} \label{eq: daynamics of sigma xpm}
\end{align}
In these equations, terms involving $a_x$, $a_p$, $b_x$, $b_p$ represent the contributions from feedback control. We note that the steady-state value of $\boldsymbol{\sigma}_{\rm c}$ does not depend on feedback gains [see Eqs.~(\ref{eq: dynamics of sigma xc})-(\ref{eq: dynamics of sigma xpc})]. Therefore, the reduction of $\braket{n}$ is solely determined by how much $\boldsymbol{\sigma}_{\mathrm{m}}$ can be reduced via feedback control.
For simplicity, let $a_x = b_p = g$, $a_p = b_x = 0$, and assume that $g$ is a constant.
Under these conditions, it can be shown that, in the limit $g \to \infty$, each component of $\boldsymbol{\sigma}_{\mathrm{m}}$ scales as $O(1/g)$.
This implies that $\boldsymbol{\sigma}_{\mathrm{m}}$ can be arbitrarily small in the limit of an infinite feedback rate.
Since $\boldsymbol{\sigma} \rightarrow \boldsymbol{\sigma}_{\mathrm{c}}$ holds in this regime, the cooling limit determined by~\eqref{eq: cooling limit} is achieved.
In this case, $\hat{\rho}_t^{e_t} = \hat{\rho}_t$ also holds
and the equality in~\eqref{eq: inequality for QC information flow and work rate} is achieved.
Moreover, since Eq.~\eqref{eq: inequality for information flow and transfer entropy} holds in the steady state with quantum Kalman filter, the equality in~\eqref{eq: steady state inequality} is also achieved. 

We now prove the asymptotic property $\boldsymbol{\sigma}_{\mathrm{m}}=O(1/g)$.
The time evolution equations for $\boldsymbol{\sigma}_{\mathrm{m}}$, as described by Eqs.~\eqref{eq: daynamics of sigma xm}, \eqref{eq: daynamics of sigma pm}, and \eqref{eq: daynamics of sigma xpm}, can be rewritten in vector form as follows:
\begin{align}\label{eq: daynamics of sigma m bector form}
\frac{d}{dt}\boldsymbol{s}_{\mathrm{m}} = \boldsymbol{A}\boldsymbol{s}_{\mathrm{m}} + \boldsymbol{v},
\end{align}
where
\begin{equation}\label{eq: sm}
\boldsymbol{s}_{\mathrm{m}} := [\sigma_{xx,{\mathrm{m}}}, \sigma_{pp,{\mathrm{m}}}, \sigma_{xp,{\mathrm{m}}}]^{\mathrm{T}}
\end{equation}
represents the variances and covariances of $\boldsymbol{\sigma}_{\mathrm{m}}$.
Here, $\boldsymbol{v} \coloneqq [8k\eta \sigma_{xx,\mathrm{c}}^2, 8k\eta \sigma_{xp,{\mathrm{c}}}^2, 8k\eta \sigma_{xx,\mathrm{c}} \sigma_{xp,\mathrm{c}}]^{\mathrm{T}}$, and the matrix $\boldsymbol{A}$ is given by
\[
\boldsymbol{A}\coloneqq\left(\begin{array}{ccc}
-2a_x-\gamma & 0  & 2\omega \\
0  & -2b_p - \gamma & -2\omega  \\
-\omega & \omega & -a_x-b_p- \gamma 
\end{array}\right).
\]
The eigenvalues of the $\boldsymbol{A}$ are given by
\begin{align}
& \lambda_1 = -(a_x + b_p + \gamma), \\
& \lambda_2 = -(a_x + b_p + \gamma) + \sqrt{(a_x - b_p)^2 - 4\omega^2}, \\
& \lambda_3 = -(a_x + b_p + \gamma) - \sqrt{(a_x - b_p)^2 - 4\omega^2}.
\end{align}
Using a matrix $\boldsymbol{P}$, $\boldsymbol{A}$ can be diagonalized as $\boldsymbol{P}^{-1} \boldsymbol{A} \boldsymbol{P} = \mathrm{diag}(\lambda_1, \lambda_2, \lambda_3)$. 
Defining $\tilde{\boldsymbol{s}} \coloneqq \boldsymbol{P} \boldsymbol{s}_{\mathrm{m}}$ and $\tilde{\boldsymbol{v}} \coloneqq \boldsymbol{P} \boldsymbol{v}$, the components of $\tilde{\boldsymbol{s}}$ satisfy: 
\begin{align} \label{eq: differential equation for tilde s}
\frac{d}{dt} \tilde{s}_i(t) = \lambda_i \tilde{s}_i(t) + \tilde{v}_i(t). 
\end{align} Assuming $\tilde{s}_i(0) = 0$, the solution is: \begin{align} \tilde{s}_i(t) = \mathrm{e}^{\lambda_i t} \int_0^t d\tau \  \mathrm{e}^{-\lambda_i \tau} \tilde{v}_i(\tau). \end{align}
Applying the triangle inequality to Eq.~\eqref{eq: differential equation for tilde s}, we obtain
\begin{align}\label{eq: inequality for tilde s}
\left| \tilde{s}_i(t) \right| \leq \frac{\left| \tilde{v}_{i,\max} \right|}{\left| \lambda_i \right|} \left| 1 - \mathrm{e}^{\lambda_i t} \right|,
\end{align}
where $|\tilde{v}_{i,\max}| = \max_{0 \leq \tau \leq t} |\tilde{v}_i(\tau)|$.
For $a_x = b_p = g$, we have $1/\lambda_i = O(1/g)$ and $\boldsymbol{P}=O(1)$, so the right-hand side of~\eqref{eq: inequality for tilde s} is $O(1/g)$.
Therefore, $\tilde{\boldsymbol{s}}(t) = O(1/g)$.
Since $\boldsymbol{s}_{\mathrm{m}} = \boldsymbol{P}^{-1} \tilde{\boldsymbol{s}}$, we conclude that $\boldsymbol{s}_{\mathrm{m}} = O(1/g)$.

\begin{figure*}[thbp]
\begin{center}
\includegraphics[width=0.95\linewidth]{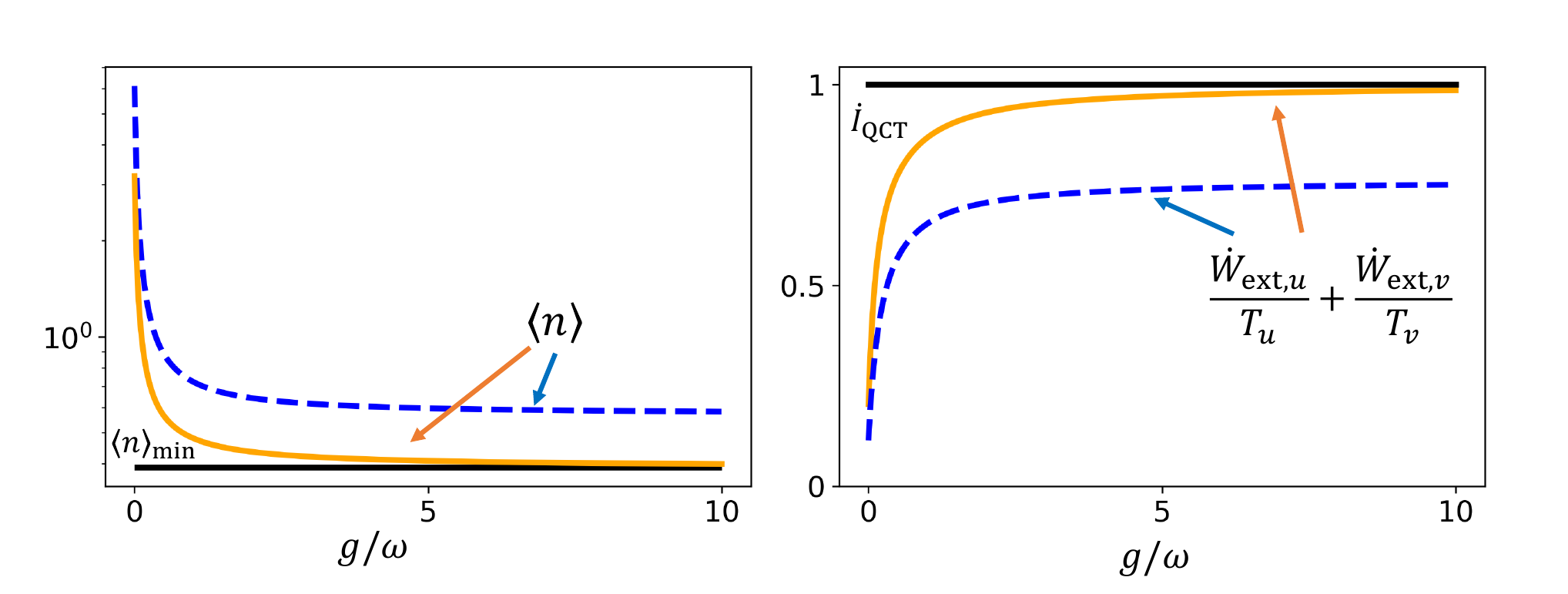}
\caption{The behavior of average occupation number $\braket{n}$ (left plot) and ${\dot{W}_{\mathrm{ext},u}}/{T_u} + {\dot{W}_{\mathrm{ext},v}}/{T_v}$ (right plot) as a function of the feedback gain $g$ in the steady state of Eq.~\eqref{eq: nanoparticle stochastic master equation}.
The feedback protocol is based on the quantum Kalman filter, where the feedback Hamiltonian is given by $H^{\hat{x},\hat{p}}_{\rm{FB}} =\hbar g(-\langle x \rangle_{\rm{c}} \hat{p} + \langle p \rangle_{\rm{c}} \hat{x})$ (orange curve).
In addition, we demonstrate another feedback Hamiltonian $H^{\hat{x}}_{\rm{FB}} =\hbar g(\langle x \rangle_{\rm{c}} + \langle p \rangle_{\rm{c}})\hat{x}$ (blue dashed curve).
In the left plot, the black line represents cooling limit $\braket{n}_{\mathrm{min}}$.
In the right plot, the black line represents the QC-transfer entropy, and the vertical axis is normalized by $\dot{I}_{\mathrm{QCT}}$.
The parameters are chosen as 
 $k/\omega = 0.18$, $\eta=0.34$, $T=292$K, and $\bar{n}\gamma/\omega =  0.0058$ ~\cite{Magrini_2021}.}
\label{fig: changing feedback gain_2}
\end{center}
\end{figure*}
We present a numerical demonstration regarding the achievement of the cooling limit.
Figure~\ref{fig: changing feedback gain_2} illustrates the behavior of the average occupation number $\langle n \rangle$ as functions of the feedback gain in the steady state described by Eq.~\eqref{eq: nanoparticle stochastic master equation}.
We compute the behavior for the two Hamiltonians $H^{\hat{x},\hat{p}}_{\rm{FB}} = g(-\langle x \rangle_{\rm{c}} \hat{p} + \langle p \rangle_{\rm{c}} \hat{x})$ and $H^{\hat{x}}_{\rm{FB}} = g(\langle x \rangle_{\rm{c}} + \langle p \rangle_{\rm{c}})\hat{x}$.
It is evident that when using $H^{\hat{x},\hat{p}}_{\rm{FB}}$, the cooling limit $\langle n \rangle_{\rm{min}}$ is asymptotically achieved as the feedback gain $g$ approaches infinity.
In contrast, for $H^{\hat{x}}_{\rm{FB}}$, the occupation number $\langle n \rangle$ asymptotically approaches a value larger than the cooling limit $\langle n \rangle_{\rm{min}}$.
In practical experiments, feedback proportional to $\hat{x}$ is commonly employed.
However, such feedback does not allow the system to reach the cooling limit $\langle n \rangle_{\rm{min}}$.
A similar trend is observed in the relationship between ${\dot{W}_{\mathrm{ext},u}}/{T_u} + {\dot{W}_{\mathrm{ext},v}}/{T_v}$ and QC-transfer entropy.

\section{Conclusion and outlook}\label{sec: conclusion and outlook}
In this study, we established a fundamental limit of quantum feedback cooling in Gaussian systems under continuous measurement, by linking work extraction, kinetic temperatures, and QC-transfer entropy~(\ref{eq: steady state inequality}).
Since the derived inequality is based on the kinetic temperatures rather than the bath temperature, it imposes a stricter constraint on cooling performance than known information-thermodynamic inequalities, and can be saturated through a sufficiently large feedback gain with the quantum Kalman filter (Fig.~\ref{fig: changing feedback gain}).
In addition, we discussed the relationship between QC-transfer entropy and yet another quantity called QC information flow, as characterized by inequality~(\ref{eq: inequality for information flow and transfer entropy}).
These results elucidate to what extent information acquired from the measurement can enhance the extraction of the work and reduce the kinetic temperatures.


Our results can be experimentally verified using current or near-future experimental technologies.
Low-frequency optomechanical systems, such as cantilever systems and nanoparticle systems, are suitable for implementing the quantum Kalman filter.
In particular, feedback cooling based on the quantum Kalman filter has been realized~\cite{Magrini_2021}, making experimental verification promising.
However, current experiments typically utilize feedback proportional only to the position operator, whereas achieving the theoretical limit requires feedback proportional to the momentum operator as well.
Realizing this optimal feedback is experimentally challenging, but is anticipated to lead to the verification of the fundamental connection between information and thermodynamics in feedback cooling.

\begin{acknowledgments}
\textit{Acknowledgments.---} 
We are grateful to Kiyotaka Aikawa and Isaac Layton for valuable discussions. 
This work was supported by JST ERATO Grant No. JPMJER2302, Japan.
K.K. and T.Y. are supported by World-leading Innovative Graduate Study Program for Materials Research, Information, and Technology (MERIT-WINGS) of the University of Tokyo. T.Y. is also supported by JSPS KAKENHI Grant No. JP23KJ0672.
K.F. is supported by JSPS KAKENHI Grant Nos. JP23K13036 and JP24H00831.
T.S. is supported by JST CREST Grant No. JPMJCR20C1. T.S. is also supported
by Institute of AI and Beyond of the University of Tokyo.
\end{acknowledgments}

\appendix

\section{Quantum information flow in general setup}
In this section, we introduce the measures of quantum information flow and derive the generalized second law, in the setup of multiple quantum measurement and feedback, which includes the continuous feedback case, mainly addressed in this manuscript, as a special case.
In Sec.~\ref{ss:QCI}, we introduce QC information flow, and derive the generalized second law. 
In Sec.~\ref{ss:QCT}, we introduce QC-transfer entropy and discuss its relationship with QC information flow. 
In Sec.~\ref{ss:cont_FB}, we demonstrate that the continuous feedback control setup discussed in the main text is a special case of the general multiple feedback setup.

\subsection{Generalized second law with QC information flow} \label{ss:QCI}

\begin{figure}[thbp]
\begin{center}
\includegraphics[width=\linewidth]{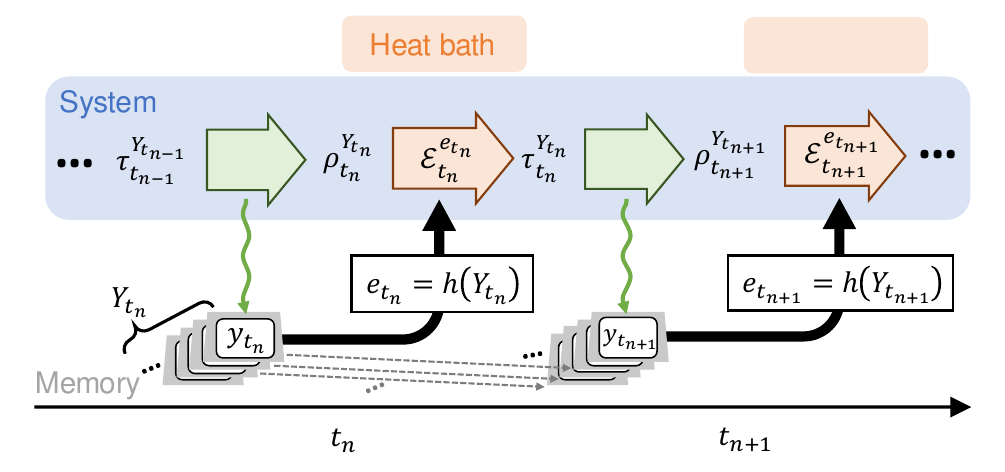}
\caption{Schematic for the multiple measurement and feedback setup. 
We consider the situation where the measurement and feedback are alternately performed.
The $n$-th measurement outcome is represented as $y_{t_n}$, and all the measurement outcomes until $y_{t_n}$ is denoted as $Y_{t_n} \coloneqq (y_{t_0},y_{t_1},\dots,y_{t_n})$. The state right before and after the \revision{$n+1$}-th measurement are denoted as $\hat{\tau}_{t_n}^{Y_{t_n}}$ and $\hat{\rho}_{t_{n+1}}^{Y_{t_{n+1}}}$, respectively, whose superscript represents the corresponding measurement outcomes. The superoperator for $n$-th feedback process is described as $\mathcal{E}_{t_n}^{e_{t_n}}$, where $e_{t_n}$ represents the estimator determined by $Y_{t_n}$.}
\label{fig:general_setup}
\end{center}
\end{figure}

We here consider the general setup of multiple measurement and feedback, as shown in Fig.~\ref{fig:general_setup}.
The process starts at the time $t=t_0$, and the $n$-th measurement and feedback is conducted during the time interval $[t_n,t_{n+1})$.
The $n$-th measurement outcome is represented as $y_{t_n}$, and all the measurement outcomes until $y_{t_n}$ is denoted as $Y_{t_n} \coloneqq (y_{t_0},y_{t_1},\dots,y_{t_n})$. 
The conditional density operators right before and after the \revision{$n+1$}-th measurement are denoted as $\hat{\tau}_{t_n}^{Y_{t_n}}$ and $\hat{\rho}_{t_{n+1}}^{Y_{t_{n+1}}}$, respectively, whose superscript represents the corresponding measurement outcomes.
The $n$-th feedback process depends on the $n$-th estimator, which is determined as a function of $Y_{t_n}$ as $e_{t_n} \coloneqq h(Y_{t_n})$, and $\mathcal{E}_{t_n}^{e_{t_n}}$ is the superoperator representing the feedback process. 
The density operator conditioned on the estimator $e_{t_n}$ is represented as $\hat{\rho}_{t_{n}}^{e_{t_{n}}}$, where the relationship 
\begin{math}
    \sum_{ h(Y_{t_n}) =e_{t_n}} P(Y_{t_n}) \hat{\rho}_{t_{n}}^{Y_{t_{n}}} = P(e_{t_n}) \hat{\rho}_{t_{n}}^{e_{t_{n}}}
\end{math}
is satisfied.
The feedback process consists of the heat-bath interaction and external driving, and therefore the superoperator can be decomposed as $\mathcal{E}_{t_n}^{e_{t_n}} = \mathcal{E}^{e_{t_n}}_{\rm drv} \circ \mathcal{E}^{e_{t_n}}_{\rm bath}$.

In such a setup, we derive the generalized second law during for the time interval spanning from the $(n_i+1)$-th to the $n_f$-th measurement and feedback process.
The quantum-classical (QC) information flow during this period can be described as 
\begin{equation}
    \label{eq:QCIS_def}
    I_{\mathrm{QCI}}^S \coloneqq  \sum_{n=n_i}^{n_f-1} \chi\left(\hat{\rho}_{t_{n+1}}: e_{t_n}\right)-\chi\left(\hat{\rho}_{t_n}: e_{t_n} \right),
\end{equation}
where $\chi (\hat{\rho}:e) \coloneqq S(\hat{\rho}) - \mathbb{E}_{e} [S(\hat{\rho}^e)]$ represents the Holevo information \cite{holevo_1973}, $S(\hat{\rho}) \coloneqq -\mathrm{tr}[\hat{\rho} \ln \hat{\rho}]$ is the von Neumann entropy, $\mathbb{E}_{X}$ represents taking the expectation value over a random variable $X$, and $\hat{\rho}_{t_{n}}\coloneqq \mathbb{E}_{e_{t_{n}}}[\hat{\rho}_{t_{n}}^{e_{t_{n}}}]$.
Since the Holevo information quantifies the correlation between the quantum system and the (classical) estimator, $-I_{\mathrm{QCI}}^S$ is the total information utilized in the multiple feedback.
The effect of measurement backaction is quantified as 
\begin{equation}
\label{eq:S_BA_gen}
    S_{\rm BA} \coloneqq \sum_{n=n_i}^{n_f-1} \mathbb{E}_{e_{t_n}} \left[ S \left(\hat{\rho}_{t_{n+1}}^{e_{t_n}}\right) -S\left(\hat{\tau}_{t_n}^{e_{t_n}}\right)\right].
\end{equation}
Since $\hat{\rho}_{t_{n+1}}^{e_{t_n}}$ is described as $\hat{\rho}_{t_{n+1}}^{e_{t_n}} = \mathcal{M}_{\mathrm{BA},t_n} (\hat{\tau}_{t_n}^{e_{t_n}})$, where $\mathcal{M}_{\mathrm{BA},t_n}$ is the superoperator representing the measurement backaction, $S_{\rm BA}$ is the accumulation of the entropy increase due to the measurement backaction.

By denoting the system Hamiltonian as $H$, the averaged heat transferred from the heat bath is defined as
\begin{align}\label{ss:heat}
    Q \coloneqq & \sum_{n=n_i}^{n_f-1} \mathbb{E}_{e_{t_n}} [Q^{e_{t_n}}]\nonumber \\ =&\sum_{n=n_i}^{n_f-1} \mathbb{E}_{e_{t_n}} \left[ \tr\left[H (\mathcal{E}^{e_{t_n}}_{\rm bath}(\hat{\rho}^{e_{t_n}}_{t_n}) -\hat{\rho}^{e_{t_n}}_{t_n}) \right] \right],
\end{align}
which quantifies the total energy increase due to the interaction with the heat bath.
The extracted work and free energy change can also be defined as 
\begin{math}
    W_{\rm ext}  \coloneqq -E(\hat{\rho}_{t_{n_f}}) + E(\hat{\rho}_{t_{n_i}}) + Q,
\end{math}
and 
\begin{math}
    \Delta F \coloneqq E(\hat{\rho}_{t_{n_f}}) - E(\hat{\rho}_{t_{n_i}}) - \beta^{-1} (S(\hat{\rho}_{t_{n_f}}) - S(\hat{\rho}_{t_{n_i}})),
\end{math}
respectively, where $E(\hat{\rho}) \coloneqq \tr[H \hat{\rho}]$ represents the internal energy of the system.

By explicitly describing the expectation value of the function $g$, $\mathbb{E}_{e} [g(e_{t_n})]$, as the sum $\sum_{e_{t_n}} P[e_{t_n}] g(e_{t_n})$, we can derive the generalized second law as follows:
\begin{equation}
\label{eq:GSL}
\begin{split}
    &\beta(W_{\rm ext} + \Delta F) + I_{\mathrm{QCI}}^S + S_{\rm BA} \\
    =& -S(\hat{\rho}_{t_{n_f}}) + S(\hat{\rho}_{t_{n_i}}) + \beta Q + I_{\mathrm{QCI}}^S + S_{\rm BA} \\
    =&  -\sum_{n={n_i}}^{n_f-1} \left\{ \mathbb{E}_{e_{t_n}} \left[ S\left(\hat{\tau}_{t_n}^{e_{t_n}}\right) - S \left(\hat{\rho}_{t_{n}}^{e_{t_n}}\right)  - \beta Q^{e_{t_n}} \right]\right\} \\
    =& -\sum_{n={n_i}}^{n_f-1} \left\{ \mathbb{E}_{e_{t_n}} \left[ S\left(\mathcal{E}^{e_{t_n}}_{\rm bath}(\hat{\rho}^{e_{t_n}}_{t_n})\right) - S \left(\hat{\rho}_{t_{n}}^{e_{t_n}}\right) \right. \right. \\
    &\quad \quad \quad \quad\left. \left.- \tr \left[ \left(\mathcal{E}^{e_{t_n}}_{\rm bath}(\hat{\rho}^{e_{t_n}}_{t_n})-\hat{\rho}^{e_{t_n}}_{t_n}\right) \ln  \hat{\rho}^{\mathrm{G}} \right]\right]\right\} \\
    =& -\sum_{n={n_i}}^{n_f-1} \left\{ \mathbb{E}_{e_{t_n}} \left[ S(\hat{\rho}^{e_{t_n}}_{t_n} \| \hat{\rho}^{\mathrm{G}}) - S(\mathcal{E}^{e_{t_n}}_{\rm bath}(\hat{\rho}^{e_{t_n}}_{t_n}) \|  \hat{\rho}^{\mathrm{G}}) \right]\right\} \\
    \leq &0,
\end{split}
\end{equation}
where $S(\hat{\rho}\|\hat{\sigma}) \coloneqq \tr[\hat{\rho} (\ln \hat{\rho} - \ln \hat{\sigma})]$ represents the quantum relative entropy, and $\hat{\rho}^{\mathrm{G}} \propto \exp(- \beta H)$ is a Gibbs state with the inverse temperature $\beta$.
The equality in the third line follows from the fact that $\hat{\tau}_{t_n}^{e_{t_n}} = \mathcal{E}^{e_{t_n}}_{\rm drv} \circ \mathcal{E}^{e_{t_n}}_{\rm bath}(\hat{\rho}^{e_{t_n}}_{t_n})$ with the superoperator $\mathcal{E}^{e_{t_n}}_{\rm drv}$ representing purely unitary driving, and therefore $S(\hat{\tau}_{t_n}^{e_{t_n}}) = S(\mathcal{E}^{e_{t_n}}_{\rm bath}(\hat{\rho}^{e_{t_n}}_{t_n}))$.
The inequality in the final line follows from the fact that $\hat{\rho}^{\mathrm{G}}$ is the fixed point of the heat-bath interaction $\mathcal{E}^{e_{t_n}}_{\rm bath}$, and the monotonicity of quantum relative entropy under a CPTP (completely positive and trace-preserving) map \cite{petz_2003}.

\subsection{QC-transfer entropy} \label{ss:QCT}

The QC-transfer entropy \cite{Yada_2022} for the time interval from the $(n_i+1)$-th to $n_f$-th process is described as 
\begin{equation}
\label{eq:QCT_def}
     I_{\mathrm{QCT}} \coloneqq \sum_{n={n_i}}^{n_f-1}  \mathbb{E}_{Y_{t_{n}}} \left[ I_{\mathrm{QC}}\left(\hat{\tau}_{t_n}^{Y_{t_n}}: y_{t_{n+1}}\right)\right],
\end{equation}
where $I_{\mathrm{QC}}$ represents the quantum-classical mutual (QC-mutual) information \cite{groenewold_1971,ozawa_1986}, defined as $I_{\mathrm{QC}}\left(\hat{\tau}: y\right) \coloneqq S(\hat{\tau}) - \mathbb{E}_{y}[S(\hat{\rho}^y)]$, with the post-selected state $\hat{\rho}^y$ for measurement outcome $y$.
The QC-mutual information $I_{\mathrm{QC}}$ quantifies the information obtained by a single measurement, and the QC-transfer entropy $I_{\mathrm{QCT}}$ is the total information transferred through the multiple measurements.
 
The QC-transfer entropy and the QC information flow satisfies the inequality
\begin{equation}
\label{eq:QCT_QCI_ineq}
    I_{\mathrm{QCT}} - \Delta\chi \geq -I_{\mathrm{QCI}}^{\mathrm{S}} -S_{\mathrm{BA}},
\end{equation}
where $\Delta \chi \coloneqq \chi (\hat{\rho}_{t_{n_f}}:Y_{t_{n_f}}) - \chi (\hat{\rho}_{t_{n_i}}:Y_{t_{n_i}}) $ is the change in correlation from $n=n_i$ to $n=n_f$.
By explicitly describing the expectation value $\mathbb{E}_{Y_{t_{n}}}[g(Y_{t_n})]$ as the sum $\sum_{Y_{t_n}} P[Y_{t_n}] g(Y_{t_n})$, we can derive this inequality as follows:
\begin{align}
    &I_{\mathrm{QCT}} - \Delta \chi + I_{\mathrm{QCI}}^{\mathrm{S}} +S_{\rm BA} \nonumber \\
    =& \sum_{n=n_i}^{n_f-1} 
    \mathbb{E}_{Y_{t_{n}}}\left[ S\left(\hat{\rho}_{t_n}^{Y_{t_n}} \| \hat{\rho}_{t_n}^{e_{t_n}}\right) - S\left(\hat{\tau}_{t_n}^{Y_{t_n}} \| \hat{\tau}_{t_n}^{e_{t_n}} \right) \right] \nonumber\\
    \geq & 0,\label{eq:QCT_QCI_ineq_derivation}
\end{align}
where the first line follows from a straightforward rearrangement of terms, and the second line follows from the equalities $\mathcal{E}_{t_n}^{e_{t_n}} (\hat{\rho}_{t_n}^{e_{t_n}}) = \hat{\tau}_{t_n}^{e_{t_n}}$ and $\mathcal{E}_{t_n}^{e_{t_n}} (\hat{\rho}_{t_n}^{Y_{t_n}}) = \hat{\tau}_{t_n}^{Y_{t_n}}$, along with the monotonicity of quantum relative entropy under a CPTP map.
As can be seen from Eq.~\eqref{eq:QCT_QCI_ineq_derivation}, a simple equality condition for Eq.~\eqref{eq:QCT_QCI_ineq} is that $\hat{\rho}_{t_n}^{e_{t_n}} = \hat{\rho}_{t_n}^{Y_{t_n}}$ holds for all $Y_{t_n}$. 
This condition implies that the feedback estimator $e_{t_n} = h(Y_{t_n})$ is optimally designed, which is known to be achieved by a Kalman filter.

\subsection{Correspondence to the continuous feedback setup considered in the main text} \label{ss:cont_FB}

We note that the definitions of thermodynamic quantities and information terms in the main text are consistent with those in Secs.~\ref{ss:QCI} and \ref{ss:QCT}. 
In particular, the consistency of the heat income $Q$, work extraction $W_{\rm ext}$, and QC information flow $I_{\mathrm{QCI}}^{\rm S}$ can be directly verified by substituting the discrete time index $t_n$ with the continuous time variable $t$ as is demonstrated in Sec.~\ref{sec: Info-thermo cooling limit derivation}.
While the correspondence is less straightforward for the measurement backaction term $S_{\mathrm{BA}}$ [Eq.~\eqref{eq: backaction entropy change} in the main text] and the QC-transfer entropy $I_{\mathrm{QCT}}$ [Eq.~\eqref{eq: QC transfer entropy} in the main text], it can still be demonstrated by neglecting terms of order $O(dt)$. 
It is comprehensive to represent Eqs.~\eqref{eq:S_BA_gen} and \eqref{eq:QCT_def} in the following forms taking the limit of $dt \to 0$:
\begin{align}
    \dot{S}_{\mathrm{BA}} &=   \mathbb{E}_{e_t} \left[ \left. \lim_{dt \to 0} \frac{S(\mathcal{M}_{\mathrm{BA},t}(\hat{\rho})) - S(\hat{\rho})}{dt}\right|_{\hat{\rho} = \hat{\tau}_{t}^{e_t}}\right], \label{eq: S_BA continuous}\\
    \dot{I}_{\mathrm{QCT}}  &= \mathbb{E}_{Y_{t}} \left[ \left. \lim_{dt \to 0} \frac{I_{\rm QC} (\hat{\rho}: dy(t)) }{dt} \right|_{\hat{\rho} = \hat{\tau}_{t}^{Y_t}}\right].\label{eq: IQCT continuous}
\end{align}
From the equalities $\hat{\tau}_{t}^{e_t} = 
\hat{\rho}_{t}^{e_t} -\frac{i}{\hbar}\left[\hat{H}_{\mathrm{HO}} + \hat{H}_{\mathrm{FB}}(t), \hat{\rho}_{t}^{e_t} \right]dt+\mathcal{L}_\mathrm{bath}[\hat{\rho}_{t}^{e_t}]dt$ and $\hat{\tau}_{t}^{Y_t} = \hat{\rho}_{t}^{Y_t} -\frac{i}{\hbar}\left[\hat{H}_{\mathrm{HO}} + \hat{H}_{\mathrm{FB}}(t), \hat{\rho}_{t}^{Y_t} \right]dt+\mathcal{L}_\mathrm{bath}[\hat{\rho}_{t}^{Y_t}]dt,$ and the fact that the
terms in the square brackets are smooth at the respective points $\hat{\rho} = \hat{\tau}_{t}^{e_t}$ and $\hat{\rho} = \hat{\tau}_{t}^{Y_t}$, we can replace $\hat{\tau}_{t}^{e_t}$ and $\hat{\tau}_{t}^{Y_t}$ in these equations with $\hat{\rho}_{t}^{e_t}$ and $\hat{\rho}_{t}^{Y_t}$, respectively, introducing only negligible additional terms of order $O(dt)$. We also note that in the main text, we use the notations $\hat{\rho}_{\rm c}(t)=\hat{\rho}_{t}^{Y_{t}}$ and $\hat{\rho}_{t,\mathrm{BA}}^{e_{t}} = \mathcal{M}_{\mathrm{BA},t}(\hat{\rho}_{t}^{e_{t}})$ for simplicity. In Sec.~\ref{sec: proof of main result}, we use the expressions~(\ref{eq: S_BA continuous}) and (\ref{eq: IQCT continuous}) to derive the main results presented in the main text.

\begin{figure}[thbp]
\begin{center}
\includegraphics[width=\linewidth]{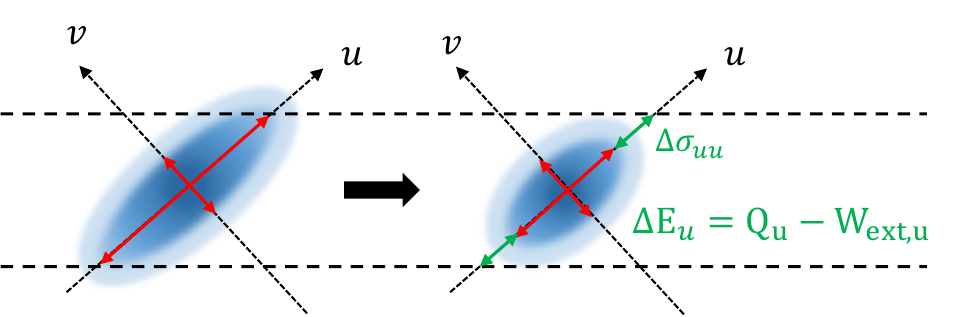}
\caption{\revision{Schematic of the change in variance along the eigenvalue directions. For simplicity, only the change along the $u$-direction is illustrated in the figure. The work extracted in the eigenvalue direction is defined from the energy change associated with the variance change due to the feedback and backaction.}}
\label{fig: sigma_u}
\end{center}
\end{figure}
\section{Proof of the results}
In this section, we derive inequalities~\eqref{eq: second law and cooling limit}, (\ref{eq: inequality for QC information flow and work rate}) and (\ref{eq: inequality for information flow and transfer entropy}) in the main text.
These inequalities are derived for a Gaussian system that obeys the following stochastic master equation~\eqref{eq: nanoparticle stochastic master equation}. We note that the dynamics of the system described by Eq.~(\ref{eq: nanoparticle stochastic master equation}) is a special case of the general setup discussed in Sec.~\ref{ss:QCI} , where the explicit form of the superoperators for the heat-bath interaction, external driving, and measurement backaction are respectively given by:
\begin{align} 
    \mathcal{E}^{e_{t}}_{\rm bath}(\hat{\rho}) &= \hat{\rho} -\frac{i}{\hbar}\left[\hat{H}_{\mathrm{HO}}, \hat{\rho} \right]dt+\mathcal{L}_\mathrm{bath}[\hat{\rho}]dt, \label{eq: bath superoperator} \\
    \mathcal{E}^{e_{t}}_{\rm drv}(\hat{\rho}) &=  \hat{\rho} -\frac{i}{\hbar}\left[\hat{H}_{\mathrm{FB}}(t), \hat{\rho} \right]dt, \label{eq: feedback superoperator} \\
    \mathcal{M}_{\mathrm{BA},t}(\hat{\rho}) &= \hat{\rho} + 2k \mathcal{D}_{\hat{x}}[\hat{\rho}] dt, \label{eq: backaction superoperator}
\end{align}
and the time interval $d t \coloneqq t_{n+1} - t_n$ is taken to the limit $d t \to 0$.
Using these superoperators, the time-evolution of the quantum state $\rho_{t}^{e_t}$ from $t$ to $t+dt$ for given $e_t$ reads:
\begin{align}\label{eq: timestep}
    \hat{\rho}_{t}^{e_{t}}\rightarrow \mathcal{E}^{e_{t}}_{\rm bath}(\hat{\rho}_{t}^{e_{t}})=:\hat{\pi}_{t}^{e_{t}} &\rightarrow \mathcal{E}^{e_{t}}_{\rm drv}(\hat{\pi}_{t}^{e_{t}}) =:\hat{\tau}_{t}^{e_{t}}  \nonumber \\
    &\rightarrow \mathcal{M}_{\mathrm{BA},t}(\hat{\tau}_{t}^{e_{t}}):= \hat{\rho}_{t+dt}^{e_t} .
\end{align}
We further assume that the feedback Hamiltonian is chosen so that, when the initial state satisfies $\langle \hat{x}\rangle(0)=\langle\hat{p}\rangle(0)=0$, the expectation values at later times satisfy $\langle \hat{x}\rangle(t)=\langle\hat{p}\rangle(t)=0$. Intuitively, such condition is satisfied when the feedback Hamiltonian is designed to reduce the mean position and momentum to the origin of the phase space. \revision{
Note that the averaged state $\hat{\rho}$ converges to a well-defined steady state under this feedback Hamiltonian, and the condition $\langle \hat{x}\rangle=\langle\hat{p}\rangle=0$ is satisfied in the steady-state as well.} On the other hand, the condition $\langle \hat{x}\rangle(t)=\langle\hat{p}\rangle(t)=0$ is not satisfied when the feedback Hamiltonian realizes a constant displacement of the position $\hat{H}_{\mathrm{FB}}(t)=a\hat{x}$, where $a$ is a constant. \revision{In this case, the  averaged state $\hat{\rho}$ also fails to reach a steady state.}

\subsection{Derivation of the inequality~\eqref{eq: inequality for QC information flow and work rate}} \label{sec: Info-thermo cooling limit derivation}
In the following, we derive~(\ref{eq: inequality for QC information flow and work rate}). We start by noting that from the definition of the heat~(\ref{eq: def of heat}), the heat rate reads
\begin{align}
    \dot{Q} := \mathbb{E}_{e_t} \left[ \lim_{dt\rightarrow 0}\frac{1}{dt} \Bigl\{ \mathrm{tr}[ (\hat{\pi}_{t}^{e_t}-\hat{\rho}_{t}^{e_{t}}) \hat{H}_{\mathrm{HO}} ] \Bigr\}  \right].
\end{align}
Similarly, the work rate reads:
\begin{align}
    \dot{W}_{\mathrm{ext}} &:=  - \mathbb{E}_{e_t} \left[ \lim_{dt\rightarrow 0} \frac{1}{dt}\Bigl\{ \mathrm{tr}\left[(\hat{\rho}_{t+dt}^{e_{t}}-\hat{\pi}_{t}^{e_{t}}) \hat{H}_{\mathrm{HO}} \right]\Bigr\} \right].
\end{align}
We then define the work rate along $\hat{u}$~(\ref{ss:u}) and $\hat{v}$~(\ref{ss:v}) as:
\begin{align}
    \dot{W}_{\mathrm{ext},u} &:= -\frac{\hbar\omega}{2} \mathbb{E}_{e_t} \left[ \lim_{dt\rightarrow 0} \frac{1}{dt}\Bigl\{\mathrm{tr}\left[(\hat{\rho}_{t+dt}^{e_{t}}-\hat{\pi}_{t}^{e_{t}}) \hat{u}^2\right]\Bigr\} \right] , \label{eq: Wextu} \\ \dot{W}_{\mathrm{ext},v}&:= - \frac{\hbar\omega}{2}\mathbb{E}_{e_t}\left[  \lim_{dt\rightarrow 0} \frac{\revision{1}}{dt}\Bigl\{ \mathrm{tr}\left[(\hat{\rho}_{t+dt}^{e_{t}}-\hat{\pi}_{t}^{e_{t}}) \hat{v}^2\right]\Bigr\} \right]. \label{eq: Wextv}
\end{align}
Here, $\dot{W}_{\mathrm{ext},u}$ quantifies the energy extracted along the $u$‐axis by feedback and measurement backaction as the variance $\sigma_{uu}$ is reduced (see Fig.~\ref{fig: sigma_u}); in this direction, the first law of thermodynamics is satisfied.
We now combine with the expression~(\ref{eq: lnrhoTuTv}) and find that the left-hand side of~(\ref{eq: inequality for QC information flow and work rate}) is expressed as 
\begin{align}\label{eq: Wext tu tv}
\frac{\dot{W}_{\mathrm{ext},u}}{T_u} + \frac{\dot{W}_{\mathrm{ext},v}}{T_v} &= \mathbb{E}_{e_t}\left[ \lim_{dt\rightarrow 0} \frac{d}{dt}\Bigl\{ \mathrm{tr}\left[ (\hat{\rho}_{t+dt}^{e_{t}}-\hat{\pi}_{t}^{e_{t}})
 \ln\hat{\rho}_{t}\right]  \Bigr\} \right] \nonumber \\
 &=  \lim_{dt\rightarrow 0} \frac{\revision{1}}{dt} \Bigl[ \mathrm{tr}\left[ (\hat{\rho}_{t+dt}-\hat{\pi}_{t})
 \ln\hat{\rho}_{t}\right]  \Bigr] \nonumber \\
 &=  \lim_{dt\rightarrow 0} \frac{\revision{1}}{dt} \Bigl[ -S(\hat{\rho}_{t+dt}) + S(\hat{\pi}_{t})   \Bigr] ,
\end{align}
where we used \revision{the condition $\langle \hat{x}\rangle(t)=\langle\hat{p}\rangle(t)=0$ and} the fact that $S(\hat{\rho}_{t+dt})=-\mathrm{tr}[\hat{\rho}_{t+dt}\ln \hat{\rho}_{t}]+O(dt^2)$, etc. 

We next discuss the QC information flow~(\ref{eq: infomation flow s}) for the time evolution given by~(\ref{eq: timestep}), which reads
\begin{align} \label{eq: QCI_S Sec2}
\dot{I}_{\mathrm{QCI}}^{\mathrm{S}}&:=\lim _{dt \rightarrow 0} \frac{\chi\left(\hat{\rho}_{t+dt}: e_{t}\right)-\chi\left(\hat{\rho}_{t}: e_{t}\right)}{d t}.
\end{align}
We further decompose Eq.~(\ref{eq: QCI_S Sec2}) into each step in~(\ref{eq: timestep}), as
\begin{align} \label{eq: bath correlation change}
\dot{I}_{\mathrm{QCI}}^{\mathrm{S}}=\dot{I}_{\mathrm{QCI,B}}^{\mathrm{S}}+\dot{I}_{\mathrm{QCI,D}}^{\mathrm{S}}+\dot{I}_{\mathrm{QCI,BA}}^{\mathrm{S}},
\end{align}
where each term is defined as:
\begin{align}
\dot{I}_{\mathrm{QCI,B}}^{\mathrm{S}}&:=\lim _{dt \rightarrow 0} \frac{\chi\left(\hat{\pi}_{t}: e_{t}\right)-\chi\left(\hat{\rho}_{t}: e_{t}\right)}{d t}, \\
    \dot{I}_{\mathrm{QCI,D}}^{\mathrm{S}}&:=\lim _{dt \rightarrow 0} \frac{\chi\left(\hat{\tau}_{t}: e_{t}\right)-\chi\left(\hat{\pi}_{t}: e_{t}\right)}{d t}, \label{eq: QCI_D} \\
    \dot{I}_{\mathrm{QCI,BA}}^{\mathrm{S}}&:= \lim _{dt \rightarrow 0} \frac{\chi\left(\hat{\rho}_{t+dt}: e_{t}\right)-\chi\left(\hat{\tau}_{t}: e_{t}\right)}{d t}.
\end{align}
Since the von Neumann entropy of the system is unchanged under the unitary time evolution~(\ref{eq: feedback superoperator}), i.e.,  $S(\hat{\pi}_{t}^{e_{t}})=S(\hat{\tau}_{t}^{e_{t}})$,  Eq.~(\ref{eq: QCI_D}) simplifies to
\begin{align} \label{eq: feedback correlation change}
\dot{I}_{\mathrm{QCI,D}}^{\mathrm{S}} = \lim _{dt \rightarrow 0} \frac{ S(\hat{\tau}_{t})-S(\hat{\pi}_{t})}{dt} . 
\end{align}
From Eqs.~\eqref{eq: bath correlation change} and~\eqref{eq: feedback correlation change}, we obtain the relation:
\begin{align} \label{eq: modyfied bath correlation change}
\dot{I}_{\mathrm{QCI}}^{\mathrm{S}}=\dot{I}_{\mathrm{QCI,B}}^{\mathrm{S}} - \dot{S}_{\mathrm{BA}} + \lim_{dt \rightarrow 0} \frac{S\left(\hat{\rho}_{t+dt}\right)-S\left(\hat{\pi}_{t}\right)}{dt}, 
\end{align}
where the entropy change of the system due to the measurement backaction $\dot{S}_{\rm BA}$ is given by Eq.~(\ref{eq: S_BA continuous}). 
We now show the following relation:
\begin{align}\label{eq: QCI_B}
\dot{I}_{\mathrm{QCI,B}}^{\mathrm{S}}\leq 0, 
\end{align}
which follows from
\begin{align}\label{eq: QCIBineq}
&\chi\left(\hat{\pi}_{t}: e_{t}\right)-\chi\left(\hat{\rho}_{t}: e_{t}\right) \nonumber \\
=& S(\mathcal{E}_{\rm bath}(\hat{\rho}_{t})) - S(\hat{\rho}_{t}) -\mathbb{E}_{e_{t}}\Bigl[ S(\mathcal{E}^{e_t}_{\rm bath}(\hat{\rho}_{t}^{e_{t}}))-S(\hat{\rho}_{t}^{e_{t}}))\Bigr] \notag \\
=&\mathbb{E}_{e_{t}}\Bigl[ S(\mathcal{E}^{e_t}_{\rm bath}(\hat{\rho}_{t}^{e_{t}})\|\mathcal{E}^{e_t}_{\rm bath}(\hat{\rho}_{t})) - S(\hat{\rho}_{t}^{e_{t}}\|\hat{\rho}_{t}) \Bigr] \notag \\
\leq &0. 
\end{align}
Here, we used the relation $\mathbb{E}_{e_t}[\mathcal{E}^{e_t}_{\rm bath}(\hat{\rho}^{e_t}_t)]=\mathcal{E}^{e_t}_{\rm bath}(\mathbb{E}_{e_t}[\hat{\rho}_{t}^{e_t}])$ which directly follows from Eq.~(\ref{eq: bath superoperator}), and the monotonicity of the quantum relative entropy and obtained the last inequality in~(\ref{eq: QCIBineq}).
By combining Eqs.~(\ref{eq: modyfied bath correlation change}), (\ref{eq: QCIBineq}), and~(\ref{eq: Wext tu tv}), we have
\begin{align} \label{eq: Ds QCIS ineq}
  \frac{\dot{W}_{\mathrm{ext},u}}{T_u} + \frac{\dot{W}_{\mathrm{ext},v}}{T_v} 
  =& -\lim_{dt \rightarrow 0} \frac{S\left(\hat{\rho}_{t+dt}\right)-S\left(\hat{\pi}_{t}\right)}{dt} \nonumber \\
  \leq & -\dot{I}_{\mathrm{QCI}}^{\mathrm{S}}-\dot{S}_{\rm BA}, 
\end{align}
which completes the proof of~(\ref{eq: inequality for QC information flow and work rate}). 

\subsection{Derivation of the inequality~\eqref{eq: inequality for information flow and transfer entropy}}\label{sec: QCT QCI derivation}
We shall now prove~(\ref{eq: inequality for information flow and transfer entropy}) by considering an infinitesimal time-step shown in Eq.~(\ref{eq: timestep}) and the monotonicity of the quantum relative entropy:
\begin{align}\label{eq:inequality for differential QCI and QCT}
    &\dot{I}_{\rm QCT}- \dot{\chi} + \dot{I}_{\rm QCI}^{\rm S}+\dot{S}_{\rm BA} \nonumber \\
 =&\lim_{dt\to 0}\mathbb{E}_{Y_t}\left[S(\hat{\rho}_t^{Y_t}\|\hat{\rho}_t^{e_t})-S(\hat{\tau}_t^{Y_t}\|\hat{\tau}_t^{e_t})\right] \geq 0,
\end{align}
where the QC-transfer entropy $\dot{I}_{\rm QCT}$ is given by Eq.~(\ref{eq: IQCT continuous}), 
and the infinitesimal variation of the Holevo information reads
\begin{align}\label{eq: def chi derivative}
    \dot{\chi} := \lim_{dt\rightarrow 0} \frac{1}{dt}\Bigl[ &S(\hat{\rho}_{t+dt})-\mathbb{E}_{Y_{t+dt}}[S(\hat{\rho}^{Y_{t+dt}}_{t+dt})]  \nonumber \\ 
    &- S(\hat{\rho}_{t})+\mathbb{E}_{Y_{t}}[S(\hat{\rho}_{t}^{Y_{t}})] \Bigr].
\end{align}
In what follows, we show that $\dot{\chi}$ vanishes in the steady-state of the Gaussian time-evolution described by Eq.~(\ref{eq: nanoparticle stochastic master equation}), thereby proving~(\ref{eq: inequality for information flow and transfer entropy}). In doing so, we first note that the von Neumann entropy of a single-mode Gaussian state can be expressed using its covariance matrix $\boldsymbol{\sigma}$ as~\cite{Olivares_2012} \begin{align}\label{eq: von neumann entropy} S(\hat{\rho}_{t})=f\left(\sqrt{\det\boldsymbol{\sigma}}\right), \end{align} where $f$ is a single-variable function defined as 
\begin{align} f(x)\coloneqq \left(x+\frac{1}{2}\right)\ln \left(x+\frac{1}{2}\right)-\left(x-\frac{1}{2}\right)\ln \left(x-\frac{1}{2}\right). \end{align}
In particular, the von Neumann entropy of the conditional state $\hat{\rho}_{t}^{Y_{t}}$ is solely determined by $\boldsymbol{\sigma}_{\rm c}(t)\coloneqq[\sigma_{ij,\mathrm{c}}(t)]_{i,j = x,p}$, where $\sigma_{xx,\mathrm{c}}(t) \coloneqq \mathrm{tr}(\hat{x}^{2}\hat{\rho}_{t}^{Y_{t}}) - [\mathrm{tr}(\hat{x}\hat{\rho}_{t}^{Y_{t}})]^2$, and similar definition applies to $\sigma_{pp,\mathrm{c}}(t)$ and $\sigma_{xp,\mathrm{c}}(t)$. For Gaussian dynamics described by Eq.~(\ref{eq: nanoparticle stochastic master equation}), we note that $\boldsymbol{\sigma}_{\rm c}$ does not depend on the measurement outcome $dy$, as their time-evolution explicitly reads:
\begin{align}
    \frac{d}{d t} \sigma_{xx,{\mathrm{c}}}&=2\omega \sigma_{xp,{\mathrm{c}}}-\gamma \sigma_{xx,{\mathrm{c}}}+\frac{\gamma}{2}(2 \bar{n}+1)-8 k\eta \sigma_{xx,{\mathrm{c}}}^2 \label{eq: dynamics of sigma xc} \\
    \frac{d}{d t} \sigma_{pp,\mathrm{c}}&=-2 \omega \sigma_{xp,{\mathrm{c}}}-\gamma \sigma_{pp,{\mathrm{c}}}+\frac{\gamma}{2}(2 \bar{n}+1)+2k-8 k\eta \sigma_{xp,{\mathrm{c}}}^2 \label{eq: dynamics of sigma pc}\\
    \frac{d}{d t} \sigma_{xp,{\mathrm{c}}}&=- \omega \sigma_{xx,{\mathrm{c}}}+\omega\sigma_{pp,{\mathrm{c}}}-\gamma \sigma_{xp,{\mathrm{c}}}-8 k\eta \sigma_{xx,{\mathrm{c}}} \sigma_{xp,{\mathrm{c}}}. \label{eq: dynamics of sigma xpc}
\end{align}
Therefore, in the steady-state, we have 
\begin{align}\label{eq: steadystate condition}
\hat{\rho}_{t}=\hat{\rho}_{t+dt}, \ \ \ \boldsymbol{\sigma}_{\rm c}(t)=\boldsymbol{\sigma}_{\rm c}(t+dt).
\end{align}
By combining Eqs.~(\ref{eq:inequality for differential QCI and QCT}), (\ref{eq: def chi derivative}) and (\ref{eq: steadystate condition}), we complete the proof of~(\ref{eq: inequality for information flow and transfer entropy}). As discussed in Sec.~\ref{ss:QCT} in the appendix, the equality in~(\ref{eq: inequality for information flow and transfer entropy}) is achieved by a quantum Kalman filter.

\subsection{Derivation of the inequality~\eqref{eq: second law and cooling limit}}
\label{sec: comparison witha the second law}

In this section, we prove inequality~\eqref{eq: second law and cooling limit} in the steady state. 
Using the fact that the free energy change vanishes in the steady state, we obtain
\begin{align}
&\frac{\dot{W}_{\mathrm{ext},u}}{T_u}
+
\frac{\dot{W}_{\mathrm{ext},v}}{T_v}
-
\frac{\dot{W}_{\mathrm{ext}}}{T} \nonumber \\
=&\lim_{dt\to0}\frac{1}{dt}\,
\mathbb{E}_{e_t}\Bigl[\mathrm{tr}\bigl(
\mathcal{E}_{\mathrm{bath}}(\hat{\rho}_t^{e_t})(\ln\hat{\rho}^{\mathrm G}-\ln\mathcal{E}_{\mathrm{bath}}(\hat{\rho}_t)) \nonumber \\
 & \qquad \qquad \qquad-\hat{\rho}_t^{e_t}(\ln\hat{\rho}^{\mathrm G}-\ln\hat{\rho}_t)\bigr)\Bigr]\notag\\
=&\lim_{dt\to0}\frac{1}{dt}\,
\mathrm{tr}\bigl(
\mathcal{E}_{\mathrm{bath}}(\hat{\rho}_t)(\ln\hat{\rho}^{\mathrm G}-\ln\mathcal{E}_{\mathrm{bath}}(\hat{\rho}_t)) \nonumber \\
& \qquad \qquad \qquad-\hat{\rho}_t(\ln\hat{\rho}^{\mathrm G}-\ln\hat{\rho}_t)\bigr)\notag\\
=&-\lim_{dt\to0}\frac{1}{dt}\Bigl(
S\bigl(\mathcal{E}_{\mathrm{bath}}(\hat{\rho}_t)\big\|\hat{\rho}^{\mathrm G}\bigr)
-S\bigl(\hat{\rho}_t\big\|\hat{\rho}^{\mathrm G}\bigr)
\Bigr)\notag\\
\ge &0,
\end{align}
where the final inequality follows from the monotonicity of the quantum relative entropy~\cite{petz_2003}.  
Note that this relation holds regardless of any ordering between
\(T\) and \(T_u, T_v\).

\subsection{Technical notes}
For a single-mode Gaussian state, the kinetic temperature introduced in this study plays the role of converting entropy into energy.
For higher-dimensional systems or non-Gaussian states, it may be possible to generalize this relationship by extending the concept of kinetic temperature.
However, a comprehensive method for defining kinetic temperatures in such scenarios remains elusive and requires further exploration. \revision{In particular, for extensions to multi-modes, one can exploit the fact that any zero-mean multi-mode Gaussian state can be regarded as a Gibbs state subjected to generalized squeezing~\cite{Weedbrook_2012_RevModPhys.84.621}; thus, one can define effective temperatures along each eigenvector direction in the same manner of this study, making such generalizations feasible.}

\revision{In this work, we employ a linear feedback Hamiltonian.  If one uses a nonlinear feedback Hamiltonian that contains quadratic or higher-order terms in $\hat{x}$ and $\hat{p}$, 
the averaged state $\hat{\rho}$ becomes non-Gaussian. 
Since the definition of kinetic temperatures relies on the Gaussianity of the state, extending the present framework to such non-Gaussian cases is an important future direction of research.}

\section{Direct Feedback}\label{sec: direct feedback}
\revision{We} consider the dynamics of direct feedback using the measurement \revision{outcome} $dy$. In other words, $dy$ itself is used as an estimate in this feedback scheme.
The time-evolution for the direct feedback is given by
\begin{align}\label{eq: feedback hamiltonian for DF}
\mathcal{E}^{e_{t}}_{\rm drv}(\hat{\rho}_{\rm c}(t)) =\hat{\rho}_{\rm c}(t) -\frac{i}{\hbar}[ -a_x  \hat{p}+b_x \hat{x}, \hat{\rho}_{\rm c}(t)] \circ dy(t-\tau_{\rm FB}),
\end{align}
where $a_{x}$ and $b_{x}$ are the feedback gains, $\circ$ denotes the Stratonovich product~\cite{Gardiner_2010}, and $\tau_{\rm FB}$ is the feedback delay time. \revision{The Stratonovich product is a stochastic differential defined with midpoint sampling, ensuring that the usual chain rule of ordinary calculus still applies.} We consider the limit of $\tau_{\rm FB}\rightarrow 0$ that corresponds to the case of Markovian feedback~\cite{Wiseman_1994_PhysRevA.49.2133, Wiseman_Milburn_2009}.  

\begin{figure*}[thbp]
\begin{center}
\includegraphics[width=0.95\linewidth]{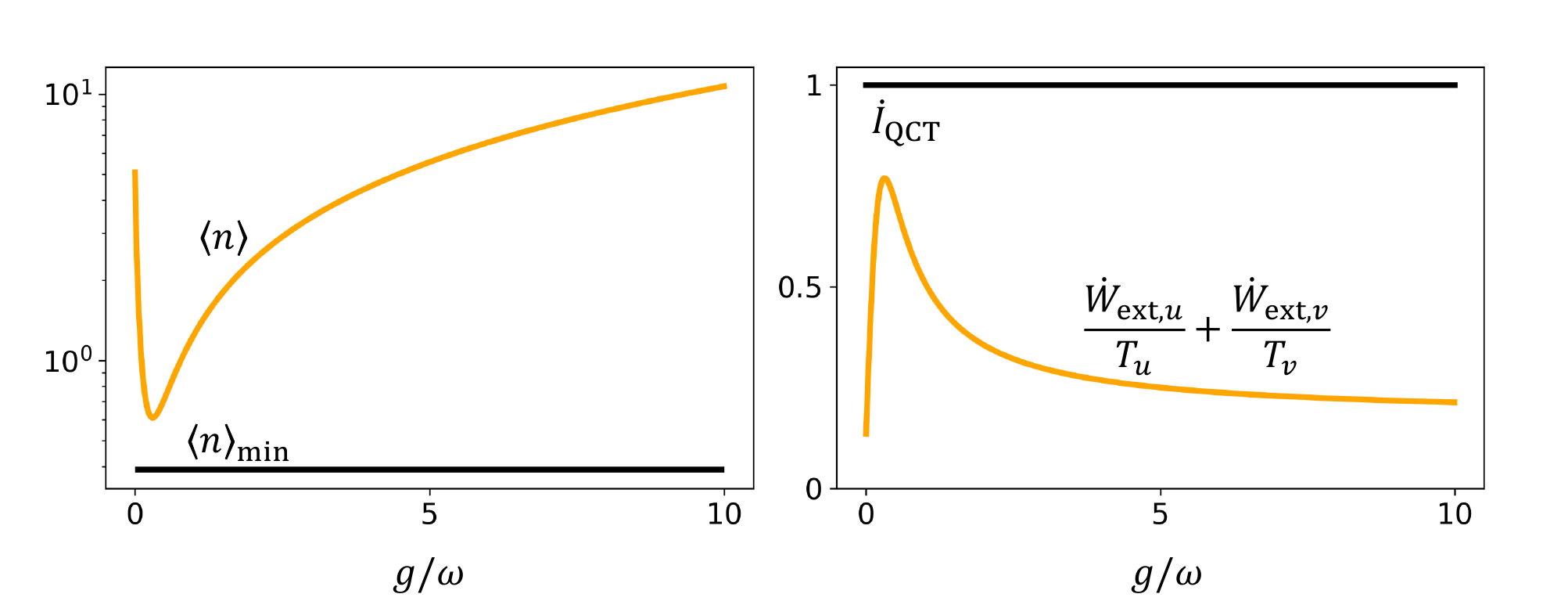}
\caption{Behavior of the steady-state average occupation number of the harmonic oscillator, $\braket{n}$ (left plot), and the quantity ${\dot{W}_{\mathrm{ext},u}}/{T_u} + {\dot{W}_{\mathrm{ext},v}}/{T_v}$ (right plot) as functions of the feedback gain $g$ under direct feedback.
The parameters are the same as in Fig.~\ref{fig: changing feedback gain}. Therefore, the values of $\braket{n}_{\rm min}$ and $\dot{I}_{\rm QCT}$ are identical to those in Fig.~\ref{fig: changing feedback gain}.
}
\label{fig: direct feedback}
\end{center}
\end{figure*}

Figure \ref{fig: direct feedback} illustrates the behavior of the steady-state average occupation number $\braket{n}$ of the harmonic oscillator and the quantity ${\dot{W}_{\mathrm{ext},u}}/{T_u} + {\dot{W}_{\mathrm{ext},v}}/{T_v}$ as functions of the feedback gain $g$ under direct feedback.
Unlike the case where the quantum Kalman filter is used, the equality in~\eqref{eq: inequality for information flow and transfer entropy} does not hold for direct feedback, and the cooling limit is not achieved.
Additionally, it is observed that the best cooling is achieved at a finite value of the feedback gain $g$, in contrast to the case of using the quantum Kalman filter, where the optimal cooling is achieved when the feedback gain becomes infinite.

\revision{
\section{Continuous position measurement}\label{sec: continious position measurement}
Here we review, following Ref.~\cite{Jacobs2006}, how the continuous position measurement used in the main text (Eq.~\eqref{eq: nanoparticle stochastic master equation} can be obtained as the infinitesimal-time limit of a discrete position measurement.
Let $\ket{\psi(t)}_{\rm c}$ denote the state conditioned on all measurement outcomes up to time $t$.  First, suppose that over a time interval $\Delta t$ the measurement is described by the operator
\begin{align}
\hat{A}(\alpha)
=\Bigl(\frac{4k\,\Delta t}{\pi}\Bigr)^{\!1/4}
\int_{-\infty}^{\infty}dx\;e^{-2k\,\Delta t\,(x-\alpha)^2}\ket{x}\bra{x}.
\end{align}
This corresponds to a measurement whose outcome $\alpha$ has a Gaussian error centered at $\alpha$.
Since $\hat{A}(\alpha)$ commutes with $\hat{x}$, it induces no backaction on the position.
However, in the presence of the harmonic potential, the position operator in the Heisenberg picture  $x(t)$ at time $t$ incorporates the effects of backaction on the momentum.
When $\Delta t$ is small, the probability density for obtaining outcome $\alpha$ is approximately
\begin{align}
P(\alpha)\approx\sqrt{\frac{4k\,\Delta t}{\pi}}
\exp\bigl[-4k\,\Delta t\,(\alpha-\braket{\hat x}_{\mathrm c})^2\bigr].
\end{align}
Hence one may write the random outcome as
\begin{align}
\alpha_s=\braket{\hat x}_{\mathrm c}
+\frac{\Delta W}{\sqrt{8k}\,\Delta t}\,.
\end{align}
Taking the $\Delta t\to0$ limit, the conditioned pure state evolves as
\begin{align}
&\ket{\psi(t+dt)}_{\mathrm c}\nonumber \\
\propto&\Bigl\{1-\bigl[k\hat x^2-4k\hat x\braket{\hat x}_{\mathrm c}\bigr]dt
+\sqrt{2k}\,\hat x\,dW\Bigr\}\ket{\psi(t)}_{\mathrm c}.
\end{align}
Converting to the density operator yields the familiar continuous measurement form
\begin{align}\label{eq: x measurement}
d\hat{\rho}_{\mathrm c}
&=-k\bigl[\hat x,\bigl[\hat x,\hat{\rho}_{\mathrm c}\bigr]\bigr]\,dt
+\sqrt{2k}\,
\bigl(\hat x\hat{\rho}_{\mathrm c}+\hat{\rho}_{\mathrm c}\hat x
-2\braket{\hat x}_{\mathrm c}\hat{\rho}_{\mathrm c}\bigr)\,dW.
\end{align}
In the main text we include an efficiency $\eta<1$ multiplying the stochastic term, reflecting the assumption that finite measurement efficiency reduces the size of the state update.  With efficiency $\eta$, the corresponding covariance dynamics become
\begin{align}
\frac{d}{dt}\sigma_{xx,{\mathrm c}}
&=-8k\eta\,\sigma_{xx,{\mathrm c}}^2,
\label{eq: dynamics of sigma xc due to x measurement}\\
\frac{d}{dt}\sigma_{pp,{\mathrm c}}
&=2k-8k\eta\,\sigma_{xp,{\mathrm c}}^2,
\label{eq: dynamics of sigma pc due to x measurement}\\
\frac{d}{dt}\sigma_{xp,{\mathrm c}}
&=-8k\eta\,\sigma_{xx,{\mathrm c}}\sigma_{xp,{\mathrm c}}.
\label{eq: dynamics of sigma xpc due to x measurement}
\end{align}
In each equation the term proportional to $-8k\eta$ corresponds to the reduction of fluctuations due to conditioning on the instantaneous position measurement.  Note that only in Eq.~\eqref{eq: dynamics of sigma pc due to x measurement} does a backaction term of size $2k$ appear, representing the momentum kick from the position measurement.
}

\section{Case of another Measurement}\label{sec: homodyne}
\revision{So far, we have considered only cases involving position measurements, which induce no backaction on the position by the measurement itself.
We here examine the effect of replacing the measurement with other measurements of the position that disturb the position and investigate how this change influences feedback cooling performance.
First, we consider the position measurement related to the operator $\hat{a}$, which is given by~\cite{Wiseman_Milburn_2009}:
\begin{align}
\label{eq: stochastic master equation for homodyne detection}
d \hat{\rho}_{\mathrm{c}}(t)=&-\frac{i}{\hbar}\left[\hat{H}_{\mathrm{HO}}+\hat{H}_{\mathrm{FB}}(t), \hat{\rho}_{\mathrm{c}}(t)\right]dt\nonumber \\
&+\mathcal{L}_\mathrm{bath}[\hat{\rho}_{\mathrm{c}}(t)]dt + 2k\mathcal{D}_{\hat{a}}[\hat{\rho}_{\mathrm{c}}(t)] dt \nonumber \\
&+ \sqrt{2k\eta}(\hat{a}\hat{\rho}_{\mathrm{c}} + \hat{\rho}_{\mathrm{c}}\hat{a}^\dagger -\sqrt{2}\braket{\hat{x}}_{\mathrm{c}}\hat{\rho}_{\mathrm{c}}) dW.
\end{align}
The measurement signal is given by $d y(t)=\braket{\hat{x}(t)}_{\mathrm{c}}dt+dW/\sqrt{8k\eta}$.
Assuming a feedback Hamiltonian of the form $\hat{H}_{\mathrm{FB}}(t)=-\left(a_x \braket{x}_{\rm{c}}+a_p\braket{p}_{\rm{c}}\right) \hat{p}+\left(b_x \braket{x}_{\rm{c}}+b_p\braket{p}_{\rm{c}}\right) \hat{x}$, the time evolution of $\braket{x}_{\mathrm{c}}$ and $\braket{p}_{\mathrm{c}}$ is governed by the following stochastic differential equations:
\begin{align}
d\braket{\hat{x}}_{\mathrm{c}}=&\left(\omega\braket{\hat{p}}_{\mathrm{c}}-a_x\braket{\hat{x}}_{\mathrm{c}}-a_p\braket{\hat{p}}_{\mathrm{c}}-\frac{\gamma}{2}\braket{\hat{x}}_{\mathrm{c}} - k\braket{\hat{x}}_{\mathrm{c}}\right)dt\nonumber \\
&+ 4\sqrt{2}k\eta\left(\sigma_{xx,\mathrm{c}}-\frac{1}{2}\right)\cdot (dy - \braket{x}_{\mathrm{c}}dt), \label{eq: ito differential equation xc with homodyne detection}\\
d\braket{\hat{p}}_{\mathrm{c}}=&\left(-\omega\braket{\hat{x}}_{\mathrm{c}}-b_x\braket{\hat{x}}_{\mathrm{c}}-b_p\braket{\hat{p}}_{\mathrm{c}}-\frac{\gamma}{2}\braket{\hat{p}}_{\mathrm{c}}- k\braket{\hat{p}}_{\mathrm{c}}\right)dt\nonumber \\
&+ 4\sqrt{2}k\eta\sigma_{xp,\mathrm{c}}\cdot (dy - \braket{x}_{\mathrm{c}}dt). \label{eq: ito differential equation pc with homodyne detection}
\end{align}
Assuming Gaussian states, the dynamics of covariance matrices $\boldsymbol{\sigma}_{\mathrm{c}}$ and $\boldsymbol{\sigma}_{\mathrm{m}}$ are given by
\begin{align}
    \frac{d}{d t} \sigma_{xx,{\mathrm{c}}}=&2\omega \sigma_{xp,{\mathrm{c}}}-\gamma \sigma_{xx,{\mathrm{c}}}+\frac{\gamma}{2}(2 \bar{n}+1) \nonumber \\
    &- 2k \sigma_{xx,{\mathrm{c}}} + k -4 k\eta \left(\sigma_{xx,{\mathrm{c}}}-\frac{1}{2}\right)^2 \label{eq: dynamics of sigma xc with homodyne detection} \\
    \frac{d}{d t} \sigma_{pp,\mathrm{c}}=&-2 \omega \sigma_{xp,{\mathrm{c}}}-\gamma \sigma_{pp,{\mathrm{c}}}+\frac{\gamma}{2}(2 \bar{n}+1)\nonumber \\
    &- 2k \sigma_{pp,{\mathrm{c}}} + k-4 k\eta \sigma_{xp,{\mathrm{c}}}^2 \label{eq: dynamics of sigma pc with homodyne detection}\\
    \frac{d}{d t} \sigma_{xp,{\mathrm{c}}}=&- \omega \sigma_{xx,{\mathrm{c}}}+\omega\sigma_{pp,{\mathrm{c}}}-\gamma \sigma_{xp,{\mathrm{c}}} \nonumber \\
    &-2k\sigma_{xp,{\mathrm{c}}}-4 k\eta \left(\sigma_{xx,{\mathrm{c}}}-\frac{1}{2}\right) \sigma_{xp,{\mathrm{c}}}, \label{eq: dynamics of sigma xpc with homodyne detection}
\end{align}
and
\begin{align}
\frac{d}{d t} \sigma_{xx,{\mathrm{m}}}=&2\omega \sigma_{xp,{\mathrm{m}}}-2 a_x \sigma_{xx,{\mathrm{m}}}-2 a_p\sigma_{xp,{\mathrm{m}}}-\gamma \sigma_{xx,{\mathrm{m}}}\nonumber \\
&- 2k \sigma_{xx,{\mathrm{m}}}+4 k\eta \left(\sigma_{xx,{\mathrm{c}}}-\frac{1}{2}\right)^2 \label{eq: daynamics of sigma xm with homodyne detection} \\
\frac{d}{d t} \sigma_{pp,{\mathrm{m}}}=&-2 \omega \sigma_{xp,{\mathrm{m}}}-2 b_x \sigma_{xp,{\mathrm{m}}}-2 b_p \sigma_{pp,{\mathrm{m}}}-\gamma \sigma_{pp,{\mathrm{m}}}\nonumber \\
&- 2k \sigma_{pp,{\mathrm{m}}}+4 k\eta \sigma_{xp,{\mathrm{c}}}^2  \label{eq: daynamics of sigma pm with homodyne detection}\\
\frac{d}{d t} \sigma_{xp,{\mathrm{m}}}=&- (\omega+b_x) \sigma_{xx,{\mathrm{m}}}+(\omega-a_p)\sigma_{pp,{\mathrm{m}}}\nonumber \\
&-(a_x+b_p+\gamma + 2k) \sigma_{xp,{\mathrm{m}}} \nonumber \\
&+4 k\eta \left(\sigma_{xx,{\mathrm{c}}}-\frac{1}{2}\right) \sigma_{xp,{\mathrm{c}}}.\label{eq: daynamics of sigma xpm with homodyne detection}
\end{align}
From Eqs.\eqref{eq: dynamics of sigma xc with homodyne detection} and \eqref{eq: dynamics of sigma xpc with homodyne detection}, we observe that $\boldsymbol{\sigma}_{\mathrm{c}}$ forms a closed set of differential equations.
This allows us to construct the quantum Kalman filter estimate using Eqs.\eqref{eq: ito differential equation xc with homodyne detection} and \eqref{eq: ito differential equation pc with homodyne detection}, in a manner analogous to the previous case.
Thus, the feedback Hamiltonian adopted here is well justified.
Moreover, an information-theoretic cooling limit can be derived using the same approach as in Sec.\ref{sec: Info-thermo cooling limit derivation}. The bound given in Eq.(10) in the main text remains valid in this case.
\begin{figure*}[thbp]
\begin{center}
\includegraphics[width=16cm]{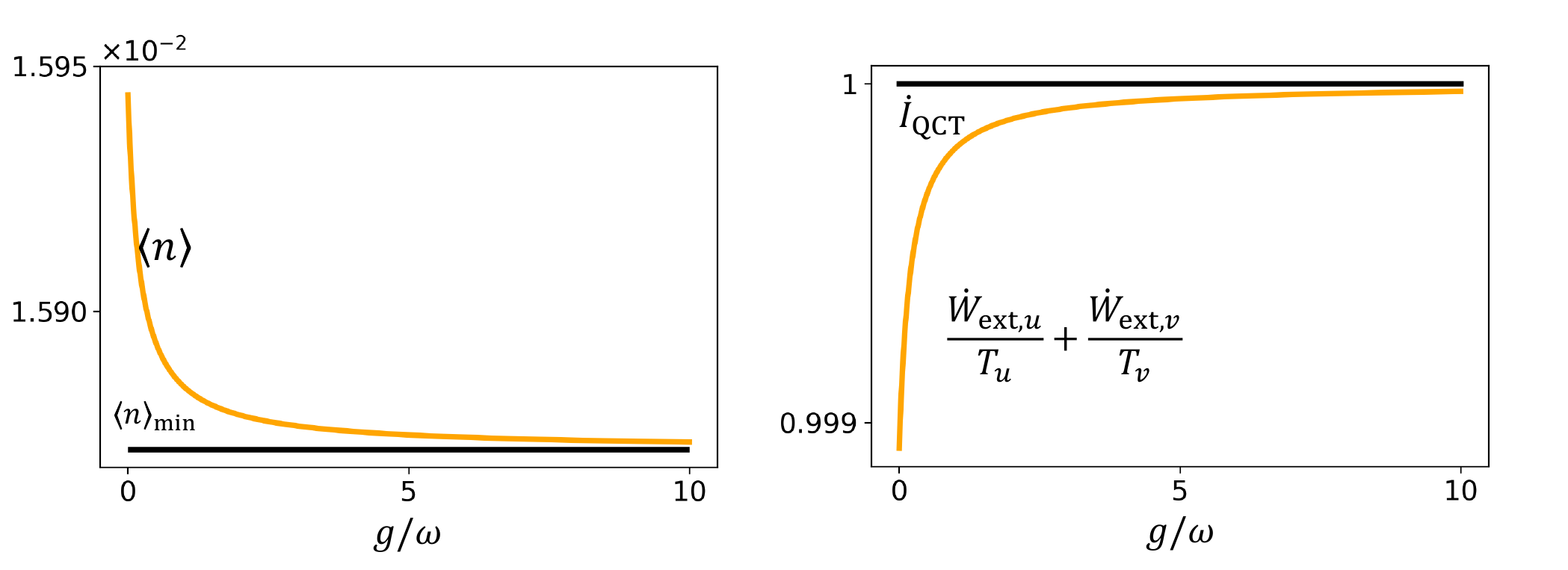}
\caption{\revision{Steady-state average occupation number $\braket{n}$ (left) and ${\dot{W}_{\rm{ext},u}}/{T_u} + {\dot{W}_{\rm{ext},v}}/{T_v}$ (right) versus feedback gain $g$, obtained from the dynamics in Eq.~\eqref{eq: stochastic master equation for homodyne detection}. The parameters are the same as in Fig.~\ref{fig: changing feedback gain}.}}
\label{fig: changing feedback gain homodyne}
\end{center}
\end{figure*}
A key distinction from our original scheme arises from the damping term $2k\mathcal{D}_{\hat{a}}[\hat{\rho}_{\mathrm{c}}(t)] dt$ in Eq.~\eqref{eq: stochastic master equation for homodyne detection}, which effectively describes coupling to a zero-temperature bath. Consequently, part of the observed cooling originates from this measurement-induced damping rather than from feedback alone, complicating direct comparisons with our original scheme.
Figure~\ref{fig: changing feedback gain homodyne} shows numerical results for the steady-state cooling performance under this dynamics. The minimum occupation number achieved is $\braket{n}_{\mathrm{min}}=0.016$, which is lower than in the original case ($\braket{n}_{\mathrm{min}}=0.39$). This improvement is likely attributable to the additional damping provided by the measurement process itself.

To compare the effect of the measurement method, it is desirable to eliminate the associated damping. To this end, we consider simultaneous detection of both the annihilation and creation operators. By construction, this setup cancels the damping terms while preserving the measurement backaction. Furthermore, we ensure that the resulting increase in $\braket{n}$ due to backaction matches that of the original case.
The stochastic master equation for this situation is:
\begin{align}
\label{eq: stochastic master equation for homodyne detection nodump}
&d \hat{\rho}_{\mathrm{c}}(t)\nonumber \\
=&-\frac{i}{\hbar}\left[\hat{H}_{\mathrm{HO}}+\hat{H}_{\mathrm{FB}}(t), \hat{\rho}_{\mathrm{c}}(t)\right]dt+\mathcal{L}_\mathrm{bath}[\hat{\rho}_{\mathrm{c}}(t)]dt \notag \\
&+ k\mathcal{D}_{\hat{a}}[\hat{\rho}_{\mathrm{c}}(t)] dt + \sqrt{\frac{k\eta}{2}}(\hat{a}\hat{\rho}_{\mathrm{c}} + \hat{\rho}_{\mathrm{c}}\hat{a}^\dagger -\sqrt{2}\braket{x}_{\mathrm{c}}\hat{\rho}_{\mathrm{c}}) dW_1 \notag \\
&+k\mathcal{D}_{\hat{a}^\dagger}[\hat{\rho}_{\mathrm{c}}(t)] dt + \sqrt{\frac{k\eta}{2}}(\hat{a^\dagger}\hat{\rho}_{\mathrm{c}} + \hat{\rho}_{\mathrm{c}}\hat{a} -\sqrt{2}\braket{x}_{\mathrm{c}}\hat{\rho}_{\mathrm{c}}) dW_2
\end{align}
Here, $dW_1$ and $dW_2$ are independent Wiener increments, and the corresponding signals are
$d y_1=\braket{\hat{x}}_{\mathrm{c}}dt+dW/2\sqrt{k\eta}$, $d y_2(t)=\braket{\hat{x}}_{\mathrm{c}}dt+dW/2\sqrt{k\eta}$, which statistically yield the same information as that for the original case, i.e., $d y=\braket{\hat{x}}_{\mathrm{c}}dt+dW/\sqrt{8k\eta}$.
The corresponding dynamics of covariance matrices $\boldsymbol{\sigma}_{\mathrm{c}}$ and $\boldsymbol{\sigma}_{\mathrm{m}}$ under Eq.~\eqref{eq: stochastic master equation for homodyne detection nodump} are given by
\begin{align}
    \frac{d}{d t} \sigma_{xx,{\mathrm{c}}}=&2\omega \sigma_{xp,{\mathrm{c}}}-\gamma \sigma_{xx,{\mathrm{c}}}+\frac{\gamma}{2}(2 \bar{n}+1) \nonumber \\
    &+ k -4 k\eta \left(\sigma_{xx,{\mathrm{c}}}^2+\frac{1}{4}\right)  \label{eq: dynamics of sigma xc with homodyne detection nodump} \end{align}
\begin{align}
    \frac{d}{d t} \sigma_{pp,\mathrm{c}}=&-2 \omega \sigma_{xp,{\mathrm{c}}}-\gamma \sigma_{pp,{\mathrm{c}}}+\frac{\gamma}{2}(2 \bar{n}+1)\nonumber \\
    &+ k-4 k\eta \sigma_{xp,{\mathrm{c}}}^2 \label{eq: dynamics of sigma pc with homodyne detection nodump}
    \end{align}
\begin{align}
    \frac{d}{d t} \sigma_{xp,{\mathrm{c}}}=&- \omega \sigma_{xx,{\mathrm{c}}}+\omega\sigma_{pp,{\mathrm{c}}}-\gamma \sigma_{xp,{\mathrm{c}}} \nonumber \\
    &-4 k\sigma_{xx,{\mathrm{c}}} \sigma_{xp,{\mathrm{c}}}. \label{eq: dynamics of sigma xpc with homodyne detection nodump}
\end{align}
and
\begin{align}
\frac{d}{d t} \sigma_{xx,{\mathrm{m}}}&=2\omega \sigma_{xp,{\mathrm{m}}}-2 a_x \sigma_{xx,{\mathrm{m}}}-2 a_p\sigma_{xp,{\mathrm{m}}}\nonumber \\
&-\gamma \sigma_{xx,{\mathrm{m}}}+4 k\eta \left(\sigma_{xx,{\mathrm{c}}}^2+\frac{1}{4}\right) \label{eq: daynamics of sigma xm with homodyne detection nodump} \\
\frac{d}{d t} \sigma_{pp,{\mathrm{m}}}&=-2 \omega \sigma_{xp,{\mathrm{m}}}-2 b_x \sigma_{xp,{\mathrm{m}}}-2 b_p \sigma_{pp,{\mathrm{m}}}\nonumber \\
&-\gamma \sigma_{pp,{\mathrm{m}}}+4 k\eta \sigma_{xp,{\mathrm{c}}}^2  \label{eq: daynamics of sigma pm with homodyne detection nodump}\\
\frac{d}{d t} \sigma_{xp,{\mathrm{m}}}&=- (\omega+b_x) \sigma_{xx,{\mathrm{m}}}+(\omega-a_p)\sigma_{pp,{\mathrm{m}}}\nonumber \\
&-(a_x+b_p+\gamma + 2k) \sigma_{xp,{\mathrm{m}}}+4 k\eta \sigma_{xx,{\mathrm{c}}}\sigma_{xp,{\mathrm{c}}}.\label{eq: daynamics of sigma xpm with homodyne detection nodump}.
\end{align}
These equations confirm that the simultaneous detection of $\hat{a}$ and $\hat{a}^\dagger$ eliminates the effective damping while retaining the measurement backaction. Compared to the our original scheme, two key differences arise: (1) the backaction now affects both position and momentum, and (2) the terms governing the measurement-induced reduction of conditional variances are different.
In particular, the final term on the right-hand side of Eq.~\eqref{eq: dynamics of sigma xpc with homodyne detection nodump} is proportional to $4k\eta\sigma_{xx,\mathrm{c}}\sigma_{xp,\mathrm{c}}$, while in the original case, the corresponding coefficient is $8k\eta\sigma_{xx,\mathrm{c}}\sigma_{xp,\mathrm{c}}$, resulting in a stronger suppression of uncertainty in the latter case. This highlights that our original scheme are more effective in reducing the conditional state variance.
\begin{figure*}[thbp]
\begin{center}
\includegraphics[width=16cm]{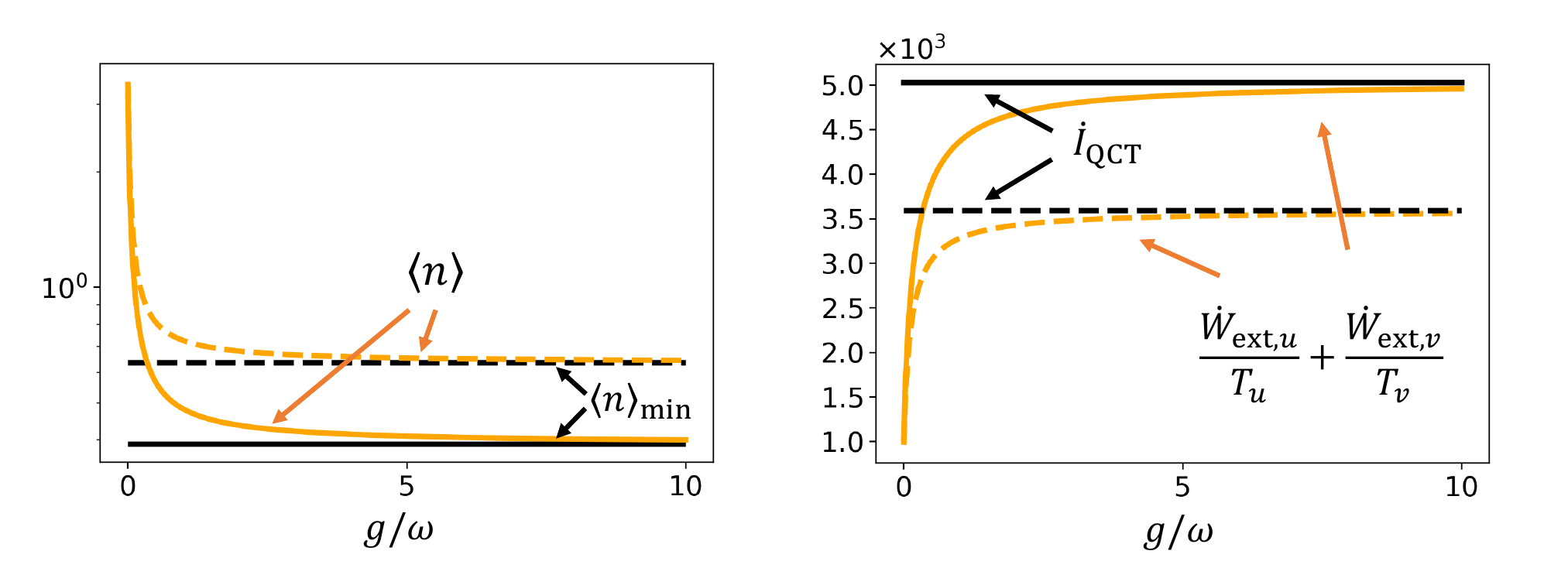}
\caption{\revision{Steady-state occupation number $\braket{n}$ (left) and ${\dot{W}_{\rm{ext},u}}/{T_u} + {\dot{W}_{\rm{ext},v}}/{T_v}$ (right) as functions of feedback gain $g$. Solid lines: measurement without position disturbance by the measurement itself [Eq.\eqref{eq: nanoparticle stochastic master equation}]. Dashed lines: measurement on $a$ and $a^\dagger$ without damping [Eq.\eqref{eq: stochastic master equation for homodyne detection nodump}]. Vertical axes are not normalized. The parameters are the same as in Fig.~\ref{fig: changing feedback gain}.}}
\label{fig: changing feedback gain homodyne nodump}
\end{center}
\end{figure*}
Figure~\ref{fig: changing feedback gain homodyne nodump} presents numerical results comparing the cooling performance under measurement with no  position disturbance by the measurement itself and measurement related to $\hat{a}$ and $\hat{a}^\dagger$ without damping. In the left panel, the minimum occupation number is lower in the original case ($\braket{n}_{\mathrm{min}}=0.39$) than in the latter case ($\braket{n}_{\mathrm{min}}=0.63$), indicating that original measurement is better in terms of feedback cooling. Since the rate of heating due to measurement backaction is the same in both scenarios, this difference in the performance is attributable to the greater reduction of conditional uncertainty achieved in the original scheme.
The right panel compares ${\dot{W}_{\rm{ext},u}}/{T_u} + {\dot{W}_{\rm{ext},v}}/{T_v}$ and $\dot{I}_{\mathrm{QCT}}$. Notably, $\dot{I}_{\mathrm{QCT}}$ differs between the two cases, with original measurements yielding higher information gain. This supports the interpretation that original measurements allow obtaining larger amount of information that can be utilized for feedback, thereby achieving lower steady-state occupation numbers.
}


\begin{thebibliography}{50}%
\makeatletter
\providecommand \@ifxundefined [1]{%
 \@ifx{#1\undefined}
}%
\providecommand \@ifnum [1]{%
 \ifnum #1\expandafter \@firstoftwo
 \else \expandafter \@secondoftwo
 \fi
}%
\providecommand \@ifx [1]{%
 \ifx #1\expandafter \@firstoftwo
 \else \expandafter \@secondoftwo
 \fi
}%
\providecommand \natexlab [1]{#1}%
\providecommand \enquote  [1]{``#1''}%
\providecommand \bibnamefont  [1]{#1}%
\providecommand \bibfnamefont [1]{#1}%
\providecommand \citenamefont [1]{#1}%
\providecommand \href@noop [0]{\@secondoftwo}%
\providecommand \href [0]{\begingroup \@sanitize@url \@href}%
\providecommand \@href[1]{\@@startlink{#1}\@@href}%
\providecommand \@@href[1]{\endgroup#1\@@endlink}%
\providecommand \@sanitize@url [0]{\catcode `\\12\catcode `\$12\catcode `\&12\catcode `\#12\catcode `\^12\catcode `\_12\catcode `\%12\relax}%
\providecommand \@@startlink[1]{}%
\providecommand \@@endlink[0]{}%
\providecommand \url  [0]{\begingroup\@sanitize@url \@url }%
\providecommand \@url [1]{\endgroup\@href {#1}{\urlprefix }}%
\providecommand \urlprefix  [0]{URL }%
\providecommand \Eprint [0]{\href }%
\providecommand \doibase [0]{https://doi.org/}%
\providecommand \selectlanguage [0]{\@gobble}%
\providecommand \bibinfo  [0]{\@secondoftwo}%
\providecommand \bibfield  [0]{\@secondoftwo}%
\providecommand \translation [1]{[#1]}%
\providecommand \BibitemOpen [0]{}%
\providecommand \bibitemStop [0]{}%
\providecommand \bibitemNoStop [0]{.\EOS\space}%
\providecommand \EOS [0]{\spacefactor3000\relax}%
\providecommand \BibitemShut  [1]{\csname bibitem#1\endcsname}%
\let\auto@bib@innerbib\@empty
\bibitem [{\citenamefont {Wiseman}\ and\ \citenamefont {Milburn}(2009)}]{Wiseman_Milburn_2009}%
  \BibitemOpen
  \bibfield  {author} {\bibinfo {author} {\bibfnamefont {H.~M.}\ \bibnamefont {Wiseman}}\ and\ \bibinfo {author} {\bibfnamefont {G.~J.}\ \bibnamefont {Milburn}},\ }\href@noop {} {\emph {\bibinfo {title} {Quantum Measurement and Control}}}\ (\bibinfo  {publisher} {Cambridge University Press},\ \bibinfo {year} {2009})\BibitemShut {NoStop}%
\bibitem [{\citenamefont {Jacobs}\ and\ \citenamefont {Steck}(2006)}]{Jacobs2006}%
  \BibitemOpen
  \bibfield  {author} {\bibinfo {author} {\bibfnamefont {K.}~\bibnamefont {Jacobs}}\ and\ \bibinfo {author} {\bibfnamefont {D.~A.}\ \bibnamefont {Steck}},\ }\bibfield  {title} {\bibinfo {title} {A straightforward introduction to continuous quantum measurement},\ }\href {https://doi.org/10.1080/00107510601101934} {\bibfield  {journal} {\bibinfo  {journal} {Contemporary Physics}\ }\textbf {\bibinfo {volume} {47}},\ \bibinfo {pages} {279} (\bibinfo {year} {2006})}\BibitemShut {NoStop}%
\bibitem [{\citenamefont {Wiseman}\ and\ \citenamefont {Milburn}(1993)}]{Wiseman1993}%
  \BibitemOpen
  \bibfield  {author} {\bibinfo {author} {\bibfnamefont {H.~M.}\ \bibnamefont {Wiseman}}\ and\ \bibinfo {author} {\bibfnamefont {G.~J.}\ \bibnamefont {Milburn}},\ }\bibfield  {title} {\bibinfo {title} {Quantum theory of optical feedback via homodyne detection},\ }\href {https://doi.org/10.1103/PhysRevLett.70.548} {\bibfield  {journal} {\bibinfo  {journal} {Phys. Rev. Lett.}\ }\textbf {\bibinfo {volume} {70}},\ \bibinfo {pages} {548} (\bibinfo {year} {1993})}\BibitemShut {NoStop}%
\bibitem [{\citenamefont {Wiseman}(1994)}]{Wiseman_1994_PhysRevA.49.2133}%
  \BibitemOpen
  \bibfield  {author} {\bibinfo {author} {\bibfnamefont {H.~M.}\ \bibnamefont {Wiseman}},\ }\bibfield  {title} {\bibinfo {title} {Quantum theory of continuous feedback},\ }\href {https://doi.org/10.1103/PhysRevA.49.2133} {\bibfield  {journal} {\bibinfo  {journal} {Phys. Rev. A}\ }\textbf {\bibinfo {volume} {49}},\ \bibinfo {pages} {2133} (\bibinfo {year} {1994})}\BibitemShut {NoStop}%
\bibitem [{\citenamefont {Doherty}\ and\ \citenamefont {Jacobs}(1999)}]{Doherty_and_Jacobs_1999}%
  \BibitemOpen
  \bibfield  {author} {\bibinfo {author} {\bibfnamefont {A.~C.}\ \bibnamefont {Doherty}}\ and\ \bibinfo {author} {\bibfnamefont {K.}~\bibnamefont {Jacobs}},\ }\bibfield  {title} {\bibinfo {title} {Feedback control of quantum systems using continuous state estimation},\ }\href {https://doi.org/10.1103/PhysRevA.60.2700} {\bibfield  {journal} {\bibinfo  {journal} {Phys. Rev. A}\ }\textbf {\bibinfo {volume} {60}},\ \bibinfo {pages} {2700} (\bibinfo {year} {1999})}\BibitemShut {NoStop}%
\bibitem [{\citenamefont {Bushev}\ \emph {et~al.}(2006)\citenamefont {Bushev}, \citenamefont {Rotter}, \citenamefont {Wilson}, \citenamefont {Dubin}, \citenamefont {Becher}, \citenamefont {Eschner}, \citenamefont {Blatt}, \citenamefont {Steixner}, \citenamefont {Rabl},\ and\ \citenamefont {Zoller}}]{PhysRevLett.96.043003}%
  \BibitemOpen
  \bibfield  {author} {\bibinfo {author} {\bibfnamefont {P.}~\bibnamefont {Bushev}}, \bibinfo {author} {\bibfnamefont {D.}~\bibnamefont {Rotter}}, \bibinfo {author} {\bibfnamefont {A.}~\bibnamefont {Wilson}}, \bibinfo {author} {\bibfnamefont {F.~m.~c.}\ \bibnamefont {Dubin}}, \bibinfo {author} {\bibfnamefont {C.}~\bibnamefont {Becher}}, \bibinfo {author} {\bibfnamefont {J.}~\bibnamefont {Eschner}}, \bibinfo {author} {\bibfnamefont {R.}~\bibnamefont {Blatt}}, \bibinfo {author} {\bibfnamefont {V.}~\bibnamefont {Steixner}}, \bibinfo {author} {\bibfnamefont {P.}~\bibnamefont {Rabl}},\ and\ \bibinfo {author} {\bibfnamefont {P.}~\bibnamefont {Zoller}},\ }\bibfield  {title} {\bibinfo {title} {Feedback cooling of a single trapped ion},\ }\href {https://doi.org/10.1103/PhysRevLett.96.043003} {\bibfield  {journal} {\bibinfo  {journal} {Phys. Rev. Lett.}\ }\textbf {\bibinfo {volume} {96}},\ \bibinfo {pages} {043003} (\bibinfo {year} {2006})}\BibitemShut {NoStop}%
\bibitem [{\citenamefont {Vijay}\ \emph {et~al.}(2012)\citenamefont {Vijay}, \citenamefont {Macklin}, \citenamefont {Slichter}, \citenamefont {Weber}, \citenamefont {Murch}, \citenamefont {Naik},\ and\ \citenamefont {Siddiqi}}]{Vijay_2012}%
  \BibitemOpen
  \bibfield  {author} {\bibinfo {author} {\bibfnamefont {R.}~\bibnamefont {Vijay}}, \bibinfo {author} {\bibfnamefont {C.}~\bibnamefont {Macklin}}, \bibinfo {author} {\bibfnamefont {D.}~\bibnamefont {Slichter}}, \bibinfo {author} {\bibfnamefont {S.}~\bibnamefont {Weber}}, \bibinfo {author} {\bibfnamefont {K.}~\bibnamefont {Murch}}, \bibinfo {author} {\bibfnamefont {R.}~\bibnamefont {Naik}},\ and\ \bibinfo {author} {\bibfnamefont {I.}~\bibnamefont {Siddiqi}},\ }\bibfield  {title} {\bibinfo {title} {Stabilizing rabi oscillations in a superconducting qubit using quantum feedback},\ }\href {https://doi.org/10.1038/nature11505} {\bibfield  {journal} {\bibinfo  {journal} {Nature}\ }\textbf {\bibinfo {volume} {490}},\ \bibinfo {pages} {77} (\bibinfo {year} {2012})}\BibitemShut {NoStop}%
\bibitem [{\citenamefont {Rist^^c3^^a8}\ \emph {et~al.}(2013)\citenamefont {Rist^^c3^^a8}, \citenamefont {Bultink}, \citenamefont {Tiggelman}, \citenamefont {Schouten}, \citenamefont {Lehnert},\ and\ \citenamefont {Dicarlo}}]{Riste_2013}%
  \BibitemOpen
  \bibfield  {author} {\bibinfo {author} {\bibfnamefont {D.}~\bibnamefont {Rist^^c3^^a8}}, \bibinfo {author} {\bibfnamefont {C.}~\bibnamefont {Bultink}}, \bibinfo {author} {\bibfnamefont {M.}~\bibnamefont {Tiggelman}}, \bibinfo {author} {\bibfnamefont {R.}~\bibnamefont {Schouten}}, \bibinfo {author} {\bibfnamefont {K.}~\bibnamefont {Lehnert}},\ and\ \bibinfo {author} {\bibfnamefont {L.}~\bibnamefont {Dicarlo}},\ }\bibfield  {title} {\bibinfo {title} {Millisecond charge-parity fluctuations and induced decoherence in a superconducting transmon qubit},\ }\href {https://doi.org/10.1038/ncomms2936} {\bibfield  {journal} {\bibinfo  {journal} {Nature Communications}\ }\textbf {\bibinfo {volume} {4}},\ \bibinfo {pages} {1913} (\bibinfo {year} {2013})}\BibitemShut {NoStop}%
\bibitem [{\citenamefont {Livingston}\ \emph {et~al.}(2022)\citenamefont {Livingston}, \citenamefont {Blok}, \citenamefont {Flurin}, \citenamefont {Dressel}, \citenamefont {Jordan},\ and\ \citenamefont {Siddiqi}}]{Livingston_2022}%
  \BibitemOpen
  \bibfield  {author} {\bibinfo {author} {\bibfnamefont {W.}~\bibnamefont {Livingston}}, \bibinfo {author} {\bibfnamefont {M.}~\bibnamefont {Blok}}, \bibinfo {author} {\bibfnamefont {E.}~\bibnamefont {Flurin}}, \bibinfo {author} {\bibfnamefont {J.}~\bibnamefont {Dressel}}, \bibinfo {author} {\bibfnamefont {A.}~\bibnamefont {Jordan}},\ and\ \bibinfo {author} {\bibfnamefont {I.}~\bibnamefont {Siddiqi}},\ }\bibfield  {title} {\bibinfo {title} {Experimental demonstration of continuous quantum error correction},\ }\href {https://doi.org/10.1038/s41467-022-29906-0} {\bibfield  {journal} {\bibinfo  {journal} {Nature Communications}\ }\textbf {\bibinfo {volume} {13}} (\bibinfo {year} {2022})}\BibitemShut {NoStop}%
\bibitem [{\citenamefont {Rossi}\ \emph {et~al.}(2018)\citenamefont {Rossi}, \citenamefont {Mason}, \citenamefont {Chen}, \citenamefont {Tsaturyan},\ and\ \citenamefont {Schliesser}}]{Rossi_2018}%
  \BibitemOpen
  \bibfield  {author} {\bibinfo {author} {\bibfnamefont {M.}~\bibnamefont {Rossi}}, \bibinfo {author} {\bibfnamefont {D.}~\bibnamefont {Mason}}, \bibinfo {author} {\bibfnamefont {J.}~\bibnamefont {Chen}}, \bibinfo {author} {\bibfnamefont {Y.}~\bibnamefont {Tsaturyan}},\ and\ \bibinfo {author} {\bibfnamefont {A.}~\bibnamefont {Schliesser}},\ }\bibfield  {title} {\bibinfo {title} {Measurement-based quantum control of mechanical motion},\ }\href {https://doi.org/10.1038/s41586-018-0643-8} {\bibfield  {journal} {\bibinfo  {journal} {Nature}\ }\textbf {\bibinfo {volume} {563}},\ \bibinfo {pages} {53} (\bibinfo {year} {2018})}\BibitemShut {NoStop}%
\bibitem [{\citenamefont {Gieseler}\ \emph {et~al.}(2012)\citenamefont {Gieseler}, \citenamefont {Deutsch}, \citenamefont {Quidant},\ and\ \citenamefont {Novotny}}]{Ginseler_2012_PhysRevLett.109.103603}%
  \BibitemOpen
  \bibfield  {author} {\bibinfo {author} {\bibfnamefont {J.}~\bibnamefont {Gieseler}}, \bibinfo {author} {\bibfnamefont {B.}~\bibnamefont {Deutsch}}, \bibinfo {author} {\bibfnamefont {R.}~\bibnamefont {Quidant}},\ and\ \bibinfo {author} {\bibfnamefont {L.}~\bibnamefont {Novotny}},\ }\bibfield  {title} {\bibinfo {title} {Subkelvin parametric feedback cooling of a laser-trapped nanoparticle},\ }\href {https://doi.org/10.1103/PhysRevLett.109.103603} {\bibfield  {journal} {\bibinfo  {journal} {Phys. Rev. Lett.}\ }\textbf {\bibinfo {volume} {109}},\ \bibinfo {pages} {103603} (\bibinfo {year} {2012})}\BibitemShut {NoStop}%
\bibitem [{\citenamefont {Magrini}\ \emph {et~al.}(2021)\citenamefont {Magrini}, \citenamefont {Rosenzweig}, \citenamefont {Bach}, \citenamefont {Deutschmann-Olek}, \citenamefont {Hofer}, \citenamefont {Hong}, \citenamefont {Kiesel}, \citenamefont {Kugi},\ and\ \citenamefont {Aspelmeyer}}]{Magrini_2021}%
  \BibitemOpen
  \bibfield  {author} {\bibinfo {author} {\bibfnamefont {L.}~\bibnamefont {Magrini}}, \bibinfo {author} {\bibfnamefont {P.}~\bibnamefont {Rosenzweig}}, \bibinfo {author} {\bibfnamefont {C.}~\bibnamefont {Bach}}, \bibinfo {author} {\bibfnamefont {A.}~\bibnamefont {Deutschmann-Olek}}, \bibinfo {author} {\bibfnamefont {S.~G.}\ \bibnamefont {Hofer}}, \bibinfo {author} {\bibfnamefont {S.}~\bibnamefont {Hong}}, \bibinfo {author} {\bibfnamefont {N.}~\bibnamefont {Kiesel}}, \bibinfo {author} {\bibfnamefont {A.}~\bibnamefont {Kugi}},\ and\ \bibinfo {author} {\bibfnamefont {M.}~\bibnamefont {Aspelmeyer}},\ }\bibfield  {title} {\bibinfo {title} {Real-time optimal quantum control of mechanical motion at room temperature},\ }\href {https://doi.org/10.1038/s41586-021-03602-3} {\bibfield  {journal} {\bibinfo  {journal} {Nature}\ }\textbf {\bibinfo {volume} {595}},\ \bibinfo {pages} {373} (\bibinfo {year} {2021})}\BibitemShut {NoStop}%
\bibitem [{\citenamefont {Kamba}\ \emph {et~al.}(2023)\citenamefont {Kamba}, \citenamefont {Shimizu},\ and\ \citenamefont {Aikawa}}]{Kamba_shimizu_aikawa_2023}%
  \BibitemOpen
  \bibfield  {author} {\bibinfo {author} {\bibfnamefont {M.}~\bibnamefont {Kamba}}, \bibinfo {author} {\bibfnamefont {R.}~\bibnamefont {Shimizu}},\ and\ \bibinfo {author} {\bibfnamefont {K.}~\bibnamefont {Aikawa}},\ }\bibfield  {title} {\bibinfo {title} {Nanoscale feedback control of six degrees of freedom of a near-sphere},\ }\href {https://doi.org/10.1038/s41467-023-43745-7} {\bibfield  {journal} {\bibinfo  {journal} {Nature Communications}\ }\textbf {\bibinfo {volume} {14}},\ \bibinfo {pages} {7943} (\bibinfo {year} {2023})}\BibitemShut {NoStop}%
\bibitem [{\citenamefont {Gonzalez-Ballestero}\ \emph {et~al.}(2021)\citenamefont {Gonzalez-Ballestero}, \citenamefont {Aspelmeyer}, \citenamefont {Novotny}, \citenamefont {Quidant},\ and\ \citenamefont {Romero-Isart}}]{doi:10.1126/science.abg3027}%
  \BibitemOpen
  \bibfield  {author} {\bibinfo {author} {\bibfnamefont {C.}~\bibnamefont {Gonzalez-Ballestero}}, \bibinfo {author} {\bibfnamefont {M.}~\bibnamefont {Aspelmeyer}}, \bibinfo {author} {\bibfnamefont {L.}~\bibnamefont {Novotny}}, \bibinfo {author} {\bibfnamefont {R.}~\bibnamefont {Quidant}},\ and\ \bibinfo {author} {\bibfnamefont {O.}~\bibnamefont {Romero-Isart}},\ }\bibfield  {title} {\bibinfo {title} {Levitodynamics: Levitation and control of microscopic objects in vacuum},\ }\href {https://doi.org/10.1126/science.abg3027} {\bibfield  {journal} {\bibinfo  {journal} {Science}\ }\textbf {\bibinfo {volume} {374}},\ \bibinfo {pages} {eabg3027} (\bibinfo {year} {2021})}\BibitemShut {NoStop}%
\bibitem [{\citenamefont {Belavkin}(1995)}]{Belavkin1995}%
  \BibitemOpen
  \bibfield  {author} {\bibinfo {author} {\bibfnamefont {V.~P.}\ \bibnamefont {Belavkin}},\ }\bibinfo {title} {Quantum filtering of markov signals with white quantum noise},\ in\ \href {https://doi.org/10.1007/978-1-4899-1391-3_37} {\emph {\bibinfo {booktitle} {Quantum Communications and Measurement}}},\ \bibinfo {editor} {edited by\ \bibinfo {editor} {\bibfnamefont {V.~P.}\ \bibnamefont {Belavkin}}, \bibinfo {editor} {\bibfnamefont {O.}~\bibnamefont {Hirota}},\ and\ \bibinfo {editor} {\bibfnamefont {R.~L.}\ \bibnamefont {Hudson}}}\ (\bibinfo  {publisher} {Springer US},\ \bibinfo {address} {Boston, MA},\ \bibinfo {year} {1995})\ pp.\ \bibinfo {pages} {381--391}\BibitemShut {NoStop}%
\bibitem [{\citenamefont {Parrondo}\ \emph {et~al.}(2015)\citenamefont {Parrondo}, \citenamefont {Horowitz},\ and\ \citenamefont {Sagawa}}]{Parrondo_Horowitz_Sagawa_2015}%
  \BibitemOpen
  \bibfield  {author} {\bibinfo {author} {\bibfnamefont {J.~M.}\ \bibnamefont {Parrondo}}, \bibinfo {author} {\bibfnamefont {J.}~\bibnamefont {Horowitz}},\ and\ \bibinfo {author} {\bibfnamefont {T.}~\bibnamefont {Sagawa}},\ }\bibfield  {title} {\bibinfo {title} {Thermodynamics of information},\ }\href {https://doi.org/10.1038/nphys3230} {\bibfield  {journal} {\bibinfo  {journal} {Nature Physics}\ }\textbf {\bibinfo {volume} {11}},\ \bibinfo {pages} {131} (\bibinfo {year} {2015})}\BibitemShut {NoStop}%
\bibitem [{\citenamefont {Sagawa}\ and\ \citenamefont {Ueda}(2008)}]{Sagawa_Ueda_2008_PhysRevLett.100.080403}%
  \BibitemOpen
  \bibfield  {author} {\bibinfo {author} {\bibfnamefont {T.}~\bibnamefont {Sagawa}}\ and\ \bibinfo {author} {\bibfnamefont {M.}~\bibnamefont {Ueda}},\ }\bibfield  {title} {\bibinfo {title} {Second law of thermodynamics with discrete quantum feedback control},\ }\href {https://doi.org/10.1103/PhysRevLett.100.080403} {\bibfield  {journal} {\bibinfo  {journal} {Phys. Rev. Lett.}\ }\textbf {\bibinfo {volume} {100}},\ \bibinfo {pages} {080403} (\bibinfo {year} {2008})}\BibitemShut {NoStop}%
\bibitem [{\citenamefont {Jacobs}(2009)}]{Jacobs_2009_PhysRevA.80.012322}%
  \BibitemOpen
  \bibfield  {author} {\bibinfo {author} {\bibfnamefont {K.}~\bibnamefont {Jacobs}},\ }\bibfield  {title} {\bibinfo {title} {Second law of thermodynamics and quantum feedback control: Maxwell's demon with weak measurements},\ }\href {https://doi.org/10.1103/PhysRevA.80.012322} {\bibfield  {journal} {\bibinfo  {journal} {Phys. Rev. A}\ }\textbf {\bibinfo {volume} {80}},\ \bibinfo {pages} {012322} (\bibinfo {year} {2009})}\BibitemShut {NoStop}%
\bibitem [{\citenamefont {Sagawa}\ and\ \citenamefont {Ueda}(2010)}]{Sagawa_Ueda_2010_PhysRevLett.104.090602}%
  \BibitemOpen
  \bibfield  {author} {\bibinfo {author} {\bibfnamefont {T.}~\bibnamefont {Sagawa}}\ and\ \bibinfo {author} {\bibfnamefont {M.}~\bibnamefont {Ueda}},\ }\bibfield  {title} {\bibinfo {title} {Generalized jarzynski equality under nonequilibrium feedback control},\ }\href {https://doi.org/10.1103/PhysRevLett.104.090602} {\bibfield  {journal} {\bibinfo  {journal} {Phys. Rev. Lett.}\ }\textbf {\bibinfo {volume} {104}},\ \bibinfo {pages} {090602} (\bibinfo {year} {2010})}\BibitemShut {NoStop}%
\bibitem [{\citenamefont {Horowitz}\ and\ \citenamefont {Vaikuntanathan}(2010)}]{Horowitz_Jordan_Vaikuntanathan_2010_PhysRevE.82.061120}%
  \BibitemOpen
  \bibfield  {author} {\bibinfo {author} {\bibfnamefont {J.~M.}\ \bibnamefont {Horowitz}}\ and\ \bibinfo {author} {\bibfnamefont {S.}~\bibnamefont {Vaikuntanathan}},\ }\bibfield  {title} {\bibinfo {title} {Nonequilibrium detailed fluctuation theorem for repeated discrete feedback},\ }\href {https://doi.org/10.1103/PhysRevE.82.061120} {\bibfield  {journal} {\bibinfo  {journal} {Phys. Rev. E}\ }\textbf {\bibinfo {volume} {82}},\ \bibinfo {pages} {061120} (\bibinfo {year} {2010})}\BibitemShut {NoStop}%
\bibitem [{\citenamefont {Funo}\ \emph {et~al.}(2013)\citenamefont {Funo}, \citenamefont {Watanabe},\ and\ \citenamefont {Ueda}}]{Funo_Watanabe_Ueda_2013_PhysRevE.88.052121}%
  \BibitemOpen
  \bibfield  {author} {\bibinfo {author} {\bibfnamefont {K.}~\bibnamefont {Funo}}, \bibinfo {author} {\bibfnamefont {Y.}~\bibnamefont {Watanabe}},\ and\ \bibinfo {author} {\bibfnamefont {M.}~\bibnamefont {Ueda}},\ }\bibfield  {title} {\bibinfo {title} {Integral quantum fluctuation theorems under measurement and feedback control},\ }\href {https://doi.org/10.1103/PhysRevE.88.052121} {\bibfield  {journal} {\bibinfo  {journal} {Phys. Rev. E}\ }\textbf {\bibinfo {volume} {88}},\ \bibinfo {pages} {052121} (\bibinfo {year} {2013})}\BibitemShut {NoStop}%
\bibitem [{\citenamefont {Gong}\ \emph {et~al.}(2016)\citenamefont {Gong}, \citenamefont {Ashida},\ and\ \citenamefont {Ueda}}]{Gong_Ashida_Ueda_PhysRevA.94.012107}%
  \BibitemOpen
  \bibfield  {author} {\bibinfo {author} {\bibfnamefont {Z.}~\bibnamefont {Gong}}, \bibinfo {author} {\bibfnamefont {Y.}~\bibnamefont {Ashida}},\ and\ \bibinfo {author} {\bibfnamefont {M.}~\bibnamefont {Ueda}},\ }\bibfield  {title} {\bibinfo {title} {Quantum-trajectory thermodynamics with discrete feedback control},\ }\href {https://doi.org/10.1103/PhysRevA.94.012107} {\bibfield  {journal} {\bibinfo  {journal} {Phys. Rev. A}\ }\textbf {\bibinfo {volume} {94}},\ \bibinfo {pages} {012107} (\bibinfo {year} {2016})}\BibitemShut {NoStop}%
\bibitem [{\citenamefont {Sagawa}\ and\ \citenamefont {Ueda}(2012)}]{Sagawa_PRE_2012}%
  \BibitemOpen
  \bibfield  {author} {\bibinfo {author} {\bibfnamefont {T.}~\bibnamefont {Sagawa}}\ and\ \bibinfo {author} {\bibfnamefont {M.}~\bibnamefont {Ueda}},\ }\bibfield  {title} {\bibinfo {title} {Nonequilibrium thermodynamics of feedback control},\ }\href {https://doi.org/10.1103/PhysRevE.85.021104} {\bibfield  {journal} {\bibinfo  {journal} {Phys. Rev. E}\ }\textbf {\bibinfo {volume} {85}},\ \bibinfo {pages} {021104} (\bibinfo {year} {2012})}\BibitemShut {NoStop}%
\bibitem [{\citenamefont {Ito}\ and\ \citenamefont {Sagawa}(2013)}]{Ito_Sagawa_2013_PhysRevLett.111.180603}%
  \BibitemOpen
  \bibfield  {author} {\bibinfo {author} {\bibfnamefont {S.}~\bibnamefont {Ito}}\ and\ \bibinfo {author} {\bibfnamefont {T.}~\bibnamefont {Sagawa}},\ }\bibfield  {title} {\bibinfo {title} {Information thermodynamics on causal networks},\ }\href {https://doi.org/10.1103/PhysRevLett.111.180603} {\bibfield  {journal} {\bibinfo  {journal} {Phys. Rev. Lett.}\ }\textbf {\bibinfo {volume} {111}},\ \bibinfo {pages} {180603} (\bibinfo {year} {2013})}\BibitemShut {NoStop}%
\bibitem [{\citenamefont {Horowitz}\ and\ \citenamefont {Esposito}(2014)}]{Horowitz_Esposito_2014_PhysRevX.4.031015}%
  \BibitemOpen
  \bibfield  {author} {\bibinfo {author} {\bibfnamefont {J.~M.}\ \bibnamefont {Horowitz}}\ and\ \bibinfo {author} {\bibfnamefont {M.}~\bibnamefont {Esposito}},\ }\bibfield  {title} {\bibinfo {title} {Thermodynamics with continuous information flow},\ }\href {https://doi.org/10.1103/PhysRevX.4.031015} {\bibfield  {journal} {\bibinfo  {journal} {Phys. Rev. X}\ }\textbf {\bibinfo {volume} {4}},\ \bibinfo {pages} {031015} (\bibinfo {year} {2014})}\BibitemShut {NoStop}%
\bibitem [{\citenamefont {Shiraishi}\ and\ \citenamefont {Sagawa}(2015)}]{Shiraishi_Sagawa_2015_PhysRevE.91.012130}%
  \BibitemOpen
  \bibfield  {author} {\bibinfo {author} {\bibfnamefont {N.}~\bibnamefont {Shiraishi}}\ and\ \bibinfo {author} {\bibfnamefont {T.}~\bibnamefont {Sagawa}},\ }\bibfield  {title} {\bibinfo {title} {Fluctuation theorem for partially masked nonequilibrium dynamics},\ }\href {https://doi.org/10.1103/PhysRevE.91.012130} {\bibfield  {journal} {\bibinfo  {journal} {Phys. Rev. E}\ }\textbf {\bibinfo {volume} {91}},\ \bibinfo {pages} {012130} (\bibinfo {year} {2015})}\BibitemShut {NoStop}%
\bibitem [{\citenamefont {Horowitz}(2015)}]{Horowitz_2015}%
  \BibitemOpen
  \bibfield  {author} {\bibinfo {author} {\bibfnamefont {J.~M.}\ \bibnamefont {Horowitz}},\ }\bibfield  {title} {\bibinfo {title} {Multipartite information flow for multiple maxwell demons},\ }\href {https://doi.org/10.1088/1742-5468/2015/03/P03006} {\bibfield  {journal} {\bibinfo  {journal} {Journal of Statistical Mechanics: Theory and Experiment}\ }\textbf {\bibinfo {volume} {2015}},\ \bibinfo {pages} {P03006} (\bibinfo {year} {2015})}\BibitemShut {NoStop}%
\bibitem [{\citenamefont {Ptaszy\ifmmode~\acute{n}\else \'{n}\fi{}ski}\ and\ \citenamefont {Esposito}(2019)}]{Krzysztof_Esposito_2019PhysRevLett.122.150603}%
  \BibitemOpen
  \bibfield  {author} {\bibinfo {author} {\bibfnamefont {K.}~\bibnamefont {Ptaszy\ifmmode~\acute{n}\else \'{n}\fi{}ski}}\ and\ \bibinfo {author} {\bibfnamefont {M.}~\bibnamefont {Esposito}},\ }\bibfield  {title} {\bibinfo {title} {Thermodynamics of quantum information flows},\ }\href {https://doi.org/10.1103/PhysRevLett.122.150603} {\bibfield  {journal} {\bibinfo  {journal} {Phys. Rev. Lett.}\ }\textbf {\bibinfo {volume} {122}},\ \bibinfo {pages} {150603} (\bibinfo {year} {2019})}\BibitemShut {NoStop}%
\bibitem [{\citenamefont {Yada}\ \emph {et~al.}(2022)\citenamefont {Yada}, \citenamefont {Yoshioka},\ and\ \citenamefont {Sagawa}}]{Yada_2022}%
  \BibitemOpen
  \bibfield  {author} {\bibinfo {author} {\bibfnamefont {T.}~\bibnamefont {Yada}}, \bibinfo {author} {\bibfnamefont {N.}~\bibnamefont {Yoshioka}},\ and\ \bibinfo {author} {\bibfnamefont {T.}~\bibnamefont {Sagawa}},\ }\bibfield  {title} {\bibinfo {title} {Quantum fluctuation theorem under quantum jumps with continuous measurement and feedback},\ }\href {https://doi.org/10.1103/PhysRevLett.128.170601} {\bibfield  {journal} {\bibinfo  {journal} {Phys. Rev. Lett.}\ }\textbf {\bibinfo {volume} {128}},\ \bibinfo {pages} {170601} (\bibinfo {year} {2022})}\BibitemShut {NoStop}%
\bibitem [{\citenamefont {Toyabe}\ \emph {et~al.}(2010)\citenamefont {Toyabe}, \citenamefont {Sagawa}, \citenamefont {Ueda}, \citenamefont {Muneyuki},\ and\ \citenamefont {Sano}}]{Toyabe_2010}%
  \BibitemOpen
  \bibfield  {author} {\bibinfo {author} {\bibfnamefont {S.}~\bibnamefont {Toyabe}}, \bibinfo {author} {\bibfnamefont {T.}~\bibnamefont {Sagawa}}, \bibinfo {author} {\bibfnamefont {M.}~\bibnamefont {Ueda}}, \bibinfo {author} {\bibfnamefont {E.}~\bibnamefont {Muneyuki}},\ and\ \bibinfo {author} {\bibfnamefont {M.}~\bibnamefont {Sano}},\ }\bibfield  {title} {\bibinfo {title} {Experimental demonstration of information-to-energy conversion and validation of the generalized jarzynski equality},\ }\href {https://doi.org/10.1038/nphys1821} {\bibfield  {journal} {\bibinfo  {journal} {Nature Physics}\ }\textbf {\bibinfo {volume} {6}},\ \bibinfo {pages} {988} (\bibinfo {year} {2010})}\BibitemShut {NoStop}%
\bibitem [{\citenamefont {B{\'e}rut}\ \emph {et~al.}(2012)\citenamefont {B{\'e}rut}, \citenamefont {Arakelyan}, \citenamefont {Petrosyan}, \citenamefont {Ciliberto}, \citenamefont {Dillenschneider},\ and\ \citenamefont {Lutz}}]{berut2012experimental}%
  \BibitemOpen
  \bibfield  {author} {\bibinfo {author} {\bibfnamefont {A.}~\bibnamefont {B{\'e}rut}}, \bibinfo {author} {\bibfnamefont {A.}~\bibnamefont {Arakelyan}}, \bibinfo {author} {\bibfnamefont {A.}~\bibnamefont {Petrosyan}}, \bibinfo {author} {\bibfnamefont {S.}~\bibnamefont {Ciliberto}}, \bibinfo {author} {\bibfnamefont {R.}~\bibnamefont {Dillenschneider}},\ and\ \bibinfo {author} {\bibfnamefont {E.}~\bibnamefont {Lutz}},\ }\bibfield  {title} {\bibinfo {title} {Experimental verification of landauer’s principle linking information and thermodynamics},\ }\href {https://doi.org/10.1038/nature10872} {\bibfield  {journal} {\bibinfo  {journal} {Nature}\ }\textbf {\bibinfo {volume} {483}},\ \bibinfo {pages} {187} (\bibinfo {year} {2012})}\BibitemShut {NoStop}%
\bibitem [{\citenamefont {Koski}\ \emph {et~al.}(2014)\citenamefont {Koski}, \citenamefont {Maisi}, \citenamefont {Pekola},\ and\ \citenamefont {Averin}}]{Koski_2014}%
  \BibitemOpen
  \bibfield  {author} {\bibinfo {author} {\bibfnamefont {J.}~\bibnamefont {Koski}}, \bibinfo {author} {\bibfnamefont {V.}~\bibnamefont {Maisi}}, \bibinfo {author} {\bibfnamefont {J.}~\bibnamefont {Pekola}},\ and\ \bibinfo {author} {\bibfnamefont {D.}~\bibnamefont {Averin}},\ }\bibfield  {title} {\bibinfo {title} {Experimental realization of a szilard engine with a single electron},\ }\href {https://doi.org/10.1073/pnas.1406966111} {\bibfield  {journal} {\bibinfo  {journal} {Proceedings of the National Academy of Sciences}\ }\textbf {\bibinfo {volume} {111}},\ \bibinfo {pages} {13786} (\bibinfo {year} {2014})}\BibitemShut {NoStop}%
\bibitem [{\citenamefont {Ribezzi-Crivellari}\ and\ \citenamefont {Ritort}(2019)}]{ribezzi2019large}%
  \BibitemOpen
  \bibfield  {author} {\bibinfo {author} {\bibfnamefont {M.}~\bibnamefont {Ribezzi-Crivellari}}\ and\ \bibinfo {author} {\bibfnamefont {F.}~\bibnamefont {Ritort}},\ }\bibfield  {title} {\bibinfo {title} {Large work extraction and the landauer limit in a continuous maxwell demon},\ }\href {https://doi.org/10.1038/s41567-019-0481-0} {\bibfield  {journal} {\bibinfo  {journal} {Nature Physics}\ }\textbf {\bibinfo {volume} {15}},\ \bibinfo {pages} {660} (\bibinfo {year} {2019})}\BibitemShut {NoStop}%
\bibitem [{\citenamefont {Debiossac}\ \emph {et~al.}(2020)\citenamefont {Debiossac}, \citenamefont {Grass}, \citenamefont {Alonso}, \citenamefont {Lutz},\ and\ \citenamefont {Kiesel}}]{debiossac_2020thermodynamics}%
  \BibitemOpen
  \bibfield  {author} {\bibinfo {author} {\bibfnamefont {M.}~\bibnamefont {Debiossac}}, \bibinfo {author} {\bibfnamefont {D.}~\bibnamefont {Grass}}, \bibinfo {author} {\bibfnamefont {J.~J.}\ \bibnamefont {Alonso}}, \bibinfo {author} {\bibfnamefont {E.}~\bibnamefont {Lutz}},\ and\ \bibinfo {author} {\bibfnamefont {N.}~\bibnamefont {Kiesel}},\ }\bibfield  {title} {\bibinfo {title} {Thermodynamics of continuous non-markovian feedback control},\ }\href {https://doi.org/10.1038/s41467-020-15148-5} {\bibfield  {journal} {\bibinfo  {journal} {Nature Communications}\ }\textbf {\bibinfo {volume} {11}},\ \bibinfo {pages} {1360} (\bibinfo {year} {2020})}\BibitemShut {NoStop}%
\bibitem [{\citenamefont {Debiossac}\ \emph {et~al.}(2022)\citenamefont {Debiossac}, \citenamefont {Rosinberg}, \citenamefont {Lutz},\ and\ \citenamefont {Kiesel}}]{debiossac2022non}%
  \BibitemOpen
  \bibfield  {author} {\bibinfo {author} {\bibfnamefont {M.}~\bibnamefont {Debiossac}}, \bibinfo {author} {\bibfnamefont {M.~L.}\ \bibnamefont {Rosinberg}}, \bibinfo {author} {\bibfnamefont {E.}~\bibnamefont {Lutz}},\ and\ \bibinfo {author} {\bibfnamefont {N.}~\bibnamefont {Kiesel}},\ }\bibfield  {title} {\bibinfo {title} {Non-markovian feedback control and acausality: An experimental study},\ }\href {https://doi.org/10.1103/PhysRevLett.128.200601} {\bibfield  {journal} {\bibinfo  {journal} {Phys. Rev. Lett.}\ }\textbf {\bibinfo {volume} {128}},\ \bibinfo {pages} {200601} (\bibinfo {year} {2022})}\BibitemShut {NoStop}%
\bibitem [{\citenamefont {Camati}\ \emph {et~al.}(2016)\citenamefont {Camati}, \citenamefont {Peterson}, \citenamefont {Batalh\~ao}, \citenamefont {Micadei}, \citenamefont {Souza}, \citenamefont {Sarthour}, \citenamefont {Oliveira},\ and\ \citenamefont {Serra}}]{Camati_2016}%
  \BibitemOpen
  \bibfield  {author} {\bibinfo {author} {\bibfnamefont {P.~A.}\ \bibnamefont {Camati}}, \bibinfo {author} {\bibfnamefont {J.~P.~S.}\ \bibnamefont {Peterson}}, \bibinfo {author} {\bibfnamefont {T.~B.}\ \bibnamefont {Batalh\~ao}}, \bibinfo {author} {\bibfnamefont {K.}~\bibnamefont {Micadei}}, \bibinfo {author} {\bibfnamefont {A.~M.}\ \bibnamefont {Souza}}, \bibinfo {author} {\bibfnamefont {R.~S.}\ \bibnamefont {Sarthour}}, \bibinfo {author} {\bibfnamefont {I.~S.}\ \bibnamefont {Oliveira}},\ and\ \bibinfo {author} {\bibfnamefont {R.~M.}\ \bibnamefont {Serra}},\ }\bibfield  {title} {\bibinfo {title} {Experimental rectification of entropy production by maxwell's demon in a quantum system},\ }\href {https://doi.org/10.1103/PhysRevLett.117.240502} {\bibfield  {journal} {\bibinfo  {journal} {Phys. Rev. Lett.}\ }\textbf {\bibinfo {volume} {117}},\ \bibinfo {pages} {240502} (\bibinfo {year} {2016})}\BibitemShut {NoStop}%
\bibitem [{\citenamefont {Cottet}\ \emph {et~al.}(2017)\citenamefont {Cottet}, \citenamefont {Jezouin}, \citenamefont {Bretheau}, \citenamefont {Campagne-Ibarcq}, \citenamefont {Ficheux}, \citenamefont {Anders}, \citenamefont {Auff^^c3^^a8ves}, \citenamefont {Azouit}, \citenamefont {Rouchon},\ and\ \citenamefont {Huard}}]{Nathanael_2017_pnas.1704827114}%
  \BibitemOpen
  \bibfield  {author} {\bibinfo {author} {\bibfnamefont {N.}~\bibnamefont {Cottet}}, \bibinfo {author} {\bibfnamefont {S.}~\bibnamefont {Jezouin}}, \bibinfo {author} {\bibfnamefont {L.}~\bibnamefont {Bretheau}}, \bibinfo {author} {\bibfnamefont {P.}~\bibnamefont {Campagne-Ibarcq}}, \bibinfo {author} {\bibfnamefont {Q.}~\bibnamefont {Ficheux}}, \bibinfo {author} {\bibfnamefont {J.}~\bibnamefont {Anders}}, \bibinfo {author} {\bibfnamefont {A.}~\bibnamefont {Auff^^c3^^a8ves}}, \bibinfo {author} {\bibfnamefont {R.}~\bibnamefont {Azouit}}, \bibinfo {author} {\bibfnamefont {P.}~\bibnamefont {Rouchon}},\ and\ \bibinfo {author} {\bibfnamefont {B.}~\bibnamefont {Huard}},\ }\bibfield  {title} {\bibinfo {title} {Observing a quantum maxwell demon at work},\ }\href {https://doi.org/10.1073/pnas.1704827114} {\bibfield  {journal} {\bibinfo  {journal} {Proceedings of the National Academy of Sciences}\ }\textbf {\bibinfo {volume} {114}},\ \bibinfo {pages} {7561} (\bibinfo {year} {2017})}\BibitemShut {NoStop}%
\bibitem [{\citenamefont {Masuyama}\ \emph {et~al.}(2018)\citenamefont {Masuyama}, \citenamefont {Funo}, \citenamefont {Murashita}, \citenamefont {Noguchi}, \citenamefont {Kono}, \citenamefont {Tabuchi}, \citenamefont {Yamazaki}, \citenamefont {Ueda},\ and\ \citenamefont {Nakamura}}]{Masuyama_2018}%
  \BibitemOpen
  \bibfield  {author} {\bibinfo {author} {\bibfnamefont {Y.}~\bibnamefont {Masuyama}}, \bibinfo {author} {\bibfnamefont {K.}~\bibnamefont {Funo}}, \bibinfo {author} {\bibfnamefont {Y.}~\bibnamefont {Murashita}}, \bibinfo {author} {\bibfnamefont {A.}~\bibnamefont {Noguchi}}, \bibinfo {author} {\bibfnamefont {S.}~\bibnamefont {Kono}}, \bibinfo {author} {\bibfnamefont {Y.}~\bibnamefont {Tabuchi}}, \bibinfo {author} {\bibfnamefont {R.}~\bibnamefont {Yamazaki}}, \bibinfo {author} {\bibfnamefont {M.}~\bibnamefont {Ueda}},\ and\ \bibinfo {author} {\bibfnamefont {Y.}~\bibnamefont {Nakamura}},\ }\bibfield  {title} {\bibinfo {title} {Information-to-work conversion by maxwell’s demon in a superconducting circuit quantum electrodynamical system},\ }\href {https://doi.org/10.1038/s41467-018-03686-y} {\bibfield  {journal} {\bibinfo  {journal} {Nature Communications}\ }\textbf {\bibinfo {volume} {9}} (\bibinfo {year} {2018})}\BibitemShut {NoStop}%
\bibitem [{\citenamefont {Naghiloo}\ \emph {et~al.}(2018)\citenamefont {Naghiloo}, \citenamefont {Alonso}, \citenamefont {Romito}, \citenamefont {Lutz},\ and\ \citenamefont {Murch}}]{Naghiloo_2018_PhysRevLett.121.030604}%
  \BibitemOpen
  \bibfield  {author} {\bibinfo {author} {\bibfnamefont {M.}~\bibnamefont {Naghiloo}}, \bibinfo {author} {\bibfnamefont {J.~J.}\ \bibnamefont {Alonso}}, \bibinfo {author} {\bibfnamefont {A.}~\bibnamefont {Romito}}, \bibinfo {author} {\bibfnamefont {E.}~\bibnamefont {Lutz}},\ and\ \bibinfo {author} {\bibfnamefont {K.~W.}\ \bibnamefont {Murch}},\ }\bibfield  {title} {\bibinfo {title} {Information gain and loss for a quantum maxwell's demon},\ }\href {https://doi.org/10.1103/PhysRevLett.121.030604} {\bibfield  {journal} {\bibinfo  {journal} {Phys. Rev. Lett.}\ }\textbf {\bibinfo {volume} {121}},\ \bibinfo {pages} {030604} (\bibinfo {year} {2018})}\BibitemShut {NoStop}%
\bibitem [{\citenamefont {Yada}\ \emph {et~al.}(2024)\citenamefont {Yada}, \citenamefont {Stas}, \citenamefont {Suleymanzade}, \citenamefont {Knall}, \citenamefont {Yoshioka}, \citenamefont {Sagawa},\ and\ \citenamefont {Lukin}}]{Yada_2024}%
  \BibitemOpen
  \bibfield  {author} {\bibinfo {author} {\bibfnamefont {T.}~\bibnamefont {Yada}}, \bibinfo {author} {\bibfnamefont {P.-J.}\ \bibnamefont {Stas}}, \bibinfo {author} {\bibfnamefont {A.}~\bibnamefont {Suleymanzade}}, \bibinfo {author} {\bibfnamefont {E.~N.}\ \bibnamefont {Knall}}, \bibinfo {author} {\bibfnamefont {N.}~\bibnamefont {Yoshioka}}, \bibinfo {author} {\bibfnamefont {T.}~\bibnamefont {Sagawa}},\ and\ \bibinfo {author} {\bibfnamefont {M.~D.}\ \bibnamefont {Lukin}},\ }\bibfield  {title} {\bibinfo {title} {Experimentally probing entropy reduction via iterative quantum information transfer},\ }\href {https://doi.org/10.48550/arXiv.2411.06709} {\bibfield  {journal} {\bibinfo  {journal} {arXiv preprint arXiv:2411.06709}\ } (\bibinfo {year} {2024})}\BibitemShut {NoStop}%
\bibitem [{\citenamefont {Horowitz}\ and\ \citenamefont {Sandberg}(2014)}]{Horowitz_and_Sandberg_2014}%
  \BibitemOpen
  \bibfield  {author} {\bibinfo {author} {\bibfnamefont {J.~M.}\ \bibnamefont {Horowitz}}\ and\ \bibinfo {author} {\bibfnamefont {H.}~\bibnamefont {Sandberg}},\ }\bibfield  {title} {\bibinfo {title} {Second-law-like inequalities with information and their interpretations},\ }\href {https://doi.org/10.1088/1367-2630/16/12/125007} {\bibfield  {journal} {\bibinfo  {journal} {New Journal of Physics}\ }\textbf {\bibinfo {volume} {16}},\ \bibinfo {pages} {125007} (\bibinfo {year} {2014})}\BibitemShut {NoStop}%
\bibitem [{\citenamefont {Sandberg}\ \emph {et~al.}(2014)\citenamefont {Sandberg}, \citenamefont {Delvenne}, \citenamefont {Newton},\ and\ \citenamefont {Mitter}}]{Sandbaerg_2014_PhysRevE.90.042119}%
  \BibitemOpen
  \bibfield  {author} {\bibinfo {author} {\bibfnamefont {H.}~\bibnamefont {Sandberg}}, \bibinfo {author} {\bibfnamefont {J.-C.}\ \bibnamefont {Delvenne}}, \bibinfo {author} {\bibfnamefont {N.~J.}\ \bibnamefont {Newton}},\ and\ \bibinfo {author} {\bibfnamefont {S.~K.}\ \bibnamefont {Mitter}},\ }\bibfield  {title} {\bibinfo {title} {Maximum work extraction and implementation costs for nonequilibrium maxwell's demons},\ }\href {https://doi.org/10.1103/PhysRevE.90.042119} {\bibfield  {journal} {\bibinfo  {journal} {Phys. Rev. E}\ }\textbf {\bibinfo {volume} {90}},\ \bibinfo {pages} {042119} (\bibinfo {year} {2014})}\BibitemShut {NoStop}%
\bibitem [{\citenamefont {Hartich}\ \emph {et~al.}(2014)\citenamefont {Hartich}, \citenamefont {Barato},\ and\ \citenamefont {Seifert}}]{Hartich_2014}%
  \BibitemOpen
  \bibfield  {author} {\bibinfo {author} {\bibfnamefont {D.}~\bibnamefont {Hartich}}, \bibinfo {author} {\bibfnamefont {A.~C.}\ \bibnamefont {Barato}},\ and\ \bibinfo {author} {\bibfnamefont {U.}~\bibnamefont {Seifert}},\ }\bibfield  {title} {\bibinfo {title} {Stochastic thermodynamics of bipartite systems: transfer entropy inequalities and a maxwell’s demon interpretation},\ }\href {https://doi.org/10.1088/1742-5468/2014/02/P02016} {\bibfield  {journal} {\bibinfo  {journal} {Journal of Statistical Mechanics: Theory and Experiment}\ }\textbf {\bibinfo {volume} {2014}},\ \bibinfo {pages} {P02016} (\bibinfo {year} {2014})}\BibitemShut {NoStop}%
\bibitem [{\citenamefont {Groenewold}(1971)}]{groenewold_1971}%
  \BibitemOpen
  \bibfield  {author} {\bibinfo {author} {\bibfnamefont {H.~J.}\ \bibnamefont {Groenewold}},\ }\bibfield  {title} {\bibinfo {title} {A problem of information gain by quantal measurements},\ }\href {https://doi.org/10.1007/BF00815357} {\bibfield  {journal} {\bibinfo  {journal} {Int. J. Theor. Phys.}\ }\textbf {\bibinfo {volume} {4}},\ \bibinfo {pages} {327} (\bibinfo {year} {1971})}\BibitemShut {NoStop}%
\bibitem [{\citenamefont {Ozawa}(1986)}]{ozawa_1986}%
  \BibitemOpen
  \bibfield  {author} {\bibinfo {author} {\bibfnamefont {M.}~\bibnamefont {Ozawa}},\ }\bibfield  {title} {\bibinfo {title} {On information gain by quantum measurements of continuous observables},\ }\href {https://doi.org/10.1063/1.527179} {\bibfield  {journal} {\bibinfo  {journal} {J. Math. Phys.}\ }\textbf {\bibinfo {volume} {27}},\ \bibinfo {pages} {759} (\bibinfo {year} {1986})}\BibitemShut {NoStop}%
\bibitem [{\citenamefont {Kim}\ \emph {et~al.}(1989)\citenamefont {Kim}, \citenamefont {de~Oliveira},\ and\ \citenamefont {Knight}}]{Kim_1989}%
  \BibitemOpen
  \bibfield  {author} {\bibinfo {author} {\bibfnamefont {M.~S.}\ \bibnamefont {Kim}}, \bibinfo {author} {\bibfnamefont {F.~A.~M.}\ \bibnamefont {de~Oliveira}},\ and\ \bibinfo {author} {\bibfnamefont {P.~L.}\ \bibnamefont {Knight}},\ }\bibfield  {title} {\bibinfo {title} {Properties of squeezed number states and squeezed thermal states},\ }\href {https://doi.org/10.1103/PhysRevA.40.2494} {\bibfield  {journal} {\bibinfo  {journal} {Phys. Rev. A}\ }\textbf {\bibinfo {volume} {40}},\ \bibinfo {pages} {2494} (\bibinfo {year} {1989})}\BibitemShut {NoStop}%
\bibitem [{\citenamefont {Olivares}(2012)}]{Olivares_2012}%
  \BibitemOpen
  \bibfield  {author} {\bibinfo {author} {\bibfnamefont {S.}~\bibnamefont {Olivares}},\ }\bibfield  {title} {\bibinfo {title} {Quantum optics in the phase space},\ }\href {https://doi.org/10.1140/epjst/e2012-01532-4} {\bibfield  {journal} {\bibinfo  {journal} {The European Physical Journal Special Topics}\ }\textbf {\bibinfo {volume} {203}},\ \bibinfo {pages} {3} (\bibinfo {year} {2012})}\BibitemShut {NoStop}%
\bibitem [{\citenamefont {Holevo}(1973)}]{holevo_1973}%
\BibitemOpen
\bibfield  {journal} {  }\bibfield  {author} {\bibinfo {author} {\bibfnamefont {A.~S.}\ \bibnamefont {Holevo}},\ }\bibfield  {title} {\bibinfo {title} {Bounds for the quantity of information transmitted by a quantum communication channel},\ }\href {https://www.mathnet.ru/eng/ppi903} {\bibfield  {journal} {\bibinfo  {journal} {Probl. Peredachi Inf.}\ }\textbf {\bibinfo {volume} {9}},\ \bibinfo {pages} {3} (\bibinfo {year} {1973})}\BibitemShut {NoStop}%
\bibitem [{\citenamefont {Matsumoto}\ and\ \citenamefont {Sagawa}(2018)}]{Matsumoto_2018}%
  \BibitemOpen
  \bibfield  {author} {\bibinfo {author} {\bibfnamefont {T.}~\bibnamefont {Matsumoto}}\ and\ \bibinfo {author} {\bibfnamefont {T.}~\bibnamefont {Sagawa}},\ }\bibfield  {title} {\bibinfo {title} {Role of sufficient statistics in stochastic thermodynamics and its implication to sensory adaptation},\ }\href {https://doi.org/10.1103/PhysRevE.97.042103} {\bibfield  {journal} {\bibinfo  {journal} {Phys. Rev. E}\ }\textbf {\bibinfo {volume} {97}},\ \bibinfo {pages} {042103} (\bibinfo {year} {2018})}\BibitemShut {NoStop}%
    \bibitem [{\citenamefont {Petz}(2003)}]{petz_2003}%
  \BibitemOpen
  \bibfield  {author} {\bibinfo {author} {\bibfnamefont {D.}~\bibnamefont {Petz}},\ }\bibfield  {title} {\bibinfo {title} {Monotonicity of quantum relative entropy revisited},\ }\href {https://doi.org/10.1142/S0129055X03001576} {\bibfield  {journal} {\bibinfo  {journal} {Rev. Math. Phys.}\ }\textbf {\bibinfo {volume} {15}},\ \bibinfo {pages} {79} (\bibinfo {year} {2003})}\BibitemShut {NoStop}%
  \bibitem [{\citenamefont {Weedbrook}\ \emph {et~al.}(2012)\citenamefont {Weedbrook}, \citenamefont {Pirandola}, \citenamefont {Garc\'{\i}a-Patr\'on}, \citenamefont {Cerf}, \citenamefont {Ralph}, \citenamefont {Shapiro},\ and\ \citenamefont {Lloyd}}]{Weedbrook_2012_RevModPhys.84.621}%
  \BibitemOpen
  \bibfield  {author} {\bibinfo {author} {\bibfnamefont {C.}~\bibnamefont {Weedbrook}}, \bibinfo {author} {\bibfnamefont {S.}~\bibnamefont {Pirandola}}, \bibinfo {author} {\bibfnamefont {R.}~\bibnamefont {Garc\'{\i}a-Patr\'on}}, \bibinfo {author} {\bibfnamefont {N.~J.}\ \bibnamefont {Cerf}}, \bibinfo {author} {\bibfnamefont {T.~C.}\ \bibnamefont {Ralph}}, \bibinfo {author} {\bibfnamefont {J.~H.}\ \bibnamefont {Shapiro}},\ and\ \bibinfo {author} {\bibfnamefont {S.}~\bibnamefont {Lloyd}},\ }\bibfield  {title} {\bibinfo {title} {Gaussian quantum information},\ }\href {https://doi.org/10.1103/RevModPhys.84.621} {\bibfield  {journal} {\bibinfo  {journal} {Rev. Mod. Phys.}\ }\textbf {\bibinfo {volume} {84}},\ \bibinfo {pages} {621} (\bibinfo {year} {2012})}\BibitemShut {NoStop}%
  \bibitem [{\citenamefont {Gardiner}(2010)}]{Gardiner_2010}%
  \BibitemOpen
  \bibfield  {author} {\bibinfo {author} {\bibfnamefont {C.}~\bibnamefont {Gardiner}},\ }\href {https://books.google.co.jp/books?id=321EuQAACAAJ} {\emph {\bibinfo {title} {Stochastic Methods: A Handbook for the Natural and Social Sciences}}},\ Springer Series in Synergetics\ (\bibinfo  {publisher} {Springer Berlin Heidelberg},\ \bibinfo {year} {2010})\BibitemShut {NoStop}%


\end{thebibliography}
\end{document}